\documentclass[10pt,journal,compsoc]{IEEEtran}

\ifCLASSOPTIONcompsoc
\usepackage[nocompress]{cite}
\else
\usepackage{cite}
\fi

\usepackage[pdftex]{graphicx}
\usepackage{epstopdf}
\graphicspath{{./}}
\DeclareGraphicsExtensions{.png,-eps-converted-to.pdf}

\ifCLASSOPTIONcompsoc
\usepackage[caption=false,font=footnotesize,labelfont=sf,textfont=sf]{subfig}
\else
\usepackage[caption=false,font=footnotesize]{subfig}
\fi

\usepackage{times}
\usepackage{amsthm}
\usepackage{amsfonts}
\usepackage{mathtools,amsmath}
\usepackage{url}
\usepackage{color, colortbl}
\usepackage{enumerate}

\usepackage{booktabs} 
\usepackage{url}

\usepackage{algorithm}
\usepackage{algpseudocode}
\usepackage{verbatim}
\usepackage{listings}
\usepackage{color}
\usepackage{textcomp}

\definecolor{codegreen}{rgb}{0,0.6,0}
\definecolor{codegray}{rgb}{0.5,0.5,0.5}
\definecolor{codepurple}{rgb}{0.58,0,0.82}
\definecolor{backcolour}{rgb}{0.95,0.95,0.92}

\lstdefinestyle{mystyle}{
	backgroundcolor=\color{backcolour},   
	commentstyle=\color{codegreen},
	keywordstyle=\color{magenta},
	numberstyle=\tiny\color{codegray},
	stringstyle=\color{codepurple},
	basicstyle=\scriptsize,
	breakatwhitespace=false,         
	breaklines=true,                 
	captionpos=b,                    
	keepspaces=true,                 
	numbers=left,                    
	numbersep=5pt,                  
	showspaces=false,                
	showstringspaces=false,
	showtabs=false,                  
	upquote=true,
	tabsize=2
}

\floatname{algorithm}{Solution Method}

\algnewcommand\algorithmicinput{\textbf{Input:}}
\algnewcommand\INPUT{\item[\algorithmicinput]}
\algnewcommand\algorithmicoutput{\textbf{Output:}}
\algnewcommand\OUTPUT{\item[\algorithmicoutput]}
\algnewcommand{\IIf}[1]{\State\algorithmicif\ #1\ \algorithmicthen}
\algnewcommand{\EndIIf}{\unskip\ \algorithmicend\ \algorithmicif}
\algrenewcommand\textproc{}

\newtheorem{definition}{Definition}

\begin{document}

\title{Modern Multicore CPUs are not Energy Proportional: Opportunity for Bi-objective Optimization for Performance and Energy}

\author{Semyon~Khokhriakov,
	Ravi~Reddy~Manumachu,
	and~Alexey~Lastovetsky
	\IEEEcompsocitemizethanks{\IEEEcompsocthanksitem S.Khokhriakhov, R. Reddy and A. Lastovetsky are with the School of Computer Science, University College Dublin, Belfield, Dublin 4, Ireland.\protect\\
		E-mail: semen.khokhriakov@ucdconnect.ie, ravi.manumachu@ucd.ie, alexey.lastovetsky@ucd.ie}
	\thanks{}}



\IEEEtitleabstractindextext{%

\begin{abstract}
Energy proportionality is the key design goal followed by architects of modern multicore CPUs. One of its implications is that optimization of an application for performance will also optimize it for energy.

In this work, we show that energy proportionality does not hold true for multicore CPUs. This finding creates the opportunity for bi-objective optimization of applications for performance and energy. We propose and study the first application-level method for bi-objective optimization of multithreaded data-parallel applications for performance and energy. The method uses two decision variables, the number of identical multithreaded kernels (threadgroups) executing the application and the number of threads in each threadgroup, so that a given workload is partitioned equally between the threadgroups.

We experimentally demonstrate the efficiency of the  method using four highly optimized multithreaded data-parallel applications, 2D fast Fourier transform based on FFTW and Intel MKL, and dense matrix-matrix multiplication using OpenBLAS and Intel MKL. Four modern multicore CPUs are used in the experiments. The experiments show that the optimization for performance alone results in the increase in dynamic energy consumption by up to 89\% and optimization for dynamic energy alone results in performance degradation by up to 49\%. By solving the bi-objective optimization problem, the method determines up to 11 globally Pareto-optimal solutions.

Finally, we propose a qualitative dynamic energy model employing performance monitoring counters (PMCs) as parameters, which we use to explain the discovered energy nonproportionality and the Pareto-optimal solutions determined by our method. The model shows that the energy nonproportionality on our experimental platforms for the two data-parallel applications is due to the activity of the data translation lookaside buffer (dTLB), which is disproportionately energy expensive.
\end{abstract}

}

\maketitle

\IEEEpeerreviewmaketitle

\IEEEraisesectionheading{\section{Introduction}\label{sec:introduction}}

Energy proportionality is the key design goal pursued by architects of modern multicore CPU platforms \cite{Barroso2007,Sen2017}. One of its implications is that optimization of an application for performance will also optimize it for energy. Modern multicore CPUs however have many inherent complexities, which are: a) Severe resource contention due to tight integration of tens of cores organized in multiple sockets with multi-level cache hierarchy and contending for shared on-chip resources such as last level cache (LLC), interconnect (For example: Intel's Quick Path Interconnect, AMD's Hyper Transport), and DRAM controllers; b) Non-uniform memory access (NUMA) where the time for memory access between a core and main memory is not uniform and where main memory is distributed between locality domains or groups called NUMA nodes; and c) Dynamic power management (DPM) of multiple power domains (CPU sockets, DRAM). 

The complexities were shown to result in complex (non-linear) functional relationships between performance and workload size and between dynamic energy and workload size for real-life data-parallel applications on modern multicore CPUs \cite{LastovetskyReddy2017,manumachu2018bi,manumachu2018bicpe}. Motivated by these research findings and based on further deep exploration, we show that energy proportionality does not hold true for multicore CPUs. This creates the opportunity for bi-objective optimization of applications for performance and energy on a single multicore CPU.

We present now an overview  of notable state-of-the-art methods solving the bi-objective optimization problem of an application for performance and energy on multicore CPU platforms. System-level methods are introduced first since they dominated the landscape. This will be followed by recent research in application-level methods. Then we describe the proposed solution method solving the bi-objective optimization problem of an application for performance and energy on a single multicore CPU.

Solution methods solving the bi-objective optimization problem for performance and energy can be broadly classified into \emph{system-level} and \emph{application-level} categories. System-level methods aim to optimize performance and energy of the environment where the applications are executed. The methods employ application-agnostic models and hardware parameters as decision variables. They are principally deployed at operating system (OS) level and therefore require changes to the OS. They do not involve any changes to the application. The methods can be further divided into the following prominent groups:
\begin{enumerate}[I.]
    \item Thread schedulers that are contention-aware and that exploit cooperative data sharing between threads \cite{Petrucci2015,Kim2017}. The goal of a scheduler is to find thread-to-core mappings to determine Pareto-optimal solutions for performance and energy. The schedulers operate at both user-level and OS-level with those at OS-level requiring changes to the OS. Thread-to-core mapping is the key decision variable. Performance monitoring counters such as LLC miss rate and LLC access rate are used for predicting the performance given a thread-to-core mapping.
    
    \item Dynamic private cache (L1 and L2) reconfiguration and shared cache (L3) partitioning strategies \cite{Wang2011,Chen2013}. The proposed solutions in this category mitigate contention for shared on-chip resources such as last level cache by physically partitioning it and therefore require substantial changes to the hardware or OS \cite{Zhuravlev2012}.
    
    \item Thermal management algorithms that place or migrate threads to not only alleviate thermal hotspots and temperature variations in a chip but also reduce energy consumption during an application execution \cite{Yang2008,Ayoub2009}. Some key strategies are dynamic power management (DPM) where idle cores are switched off, Dynamic Voltage and Frequency Scaling (DVFS), which throttles the frequencies of the cores based on their utilization, sand migration of threads from hot cores to the colder cores.
    
    \item Asymmetry-aware schedulers that exploit the asymmetry between sets of cores in a multicore platform to find thread-to-core mappings that provide Pareto-optimal solutions for performance and energy \cite{Li2007,Humenay2007}. Asymmetry can be explicit with fast and slow cores or implicit due to non-uniform frequency scaling between different cores or performance differences introduced by manufacturing variations. The key decision variables employed here are thread-to-core mapping and DVFS. Typical strategy is to map the most power-intensive threads to less power-hungry cores and then apply DVFS to the cores to ensure all threads complete at the same time whilst satisfying a power budget constraint.
\end{enumerate}

In the second category, solution methods optimize applications rather than the executing environment. The methods use application-level decision variables and predictive models for performance and energy consumption of applications to solve the bi-objective optimization problem. The dominant decision variables include the number of threads, loop tile size, workload distribution, etc. Following the principle of energy proportionality, a dominant class of such solution methods aim to achieve optimal energy reduction by optimizing for performance alone. Definitive examples are scientific routines offered by vendor-specific software packages that are extensively optimized for performance. For example, Intel Math Kernel Library \cite{IntelMKL} provides extensively optimized multithreaded basic linear algebra subprograms (BLAS) and 1D, 2D, and 3D fast Fourier transform (FFT) routines for Intel processors. Open source packages such as \cite{OpenBLAS,FFTW,ZZGemmOOC} offer the same interface functions but contain portable optimizations and may exhibit better average performance than a heavily optimized vendor package \cite{khaleghzadeh2018out,khokhriakov2018performance}. The optimized routines in these software packages allow employment of one key decision variable, which is the number of threads. A given workload is load-balanced between the threads. \textit{In this work, we show that the optimal number of threads (and consequently load-balanced workload distribution) maximizing the performance does not necessarily minimize the energy consumption of multicore CPUs.}

State-of-the-art research works on application-level optimization methods \cite{LastovetskyReddy2017,manumachu2018bi,manumachu2018bicpe} demonstrate that due to the aforementioned design complexities of modern multicore CPU platforms, the functional relationships between performance and workload size and between dynamic energy and workload size for real-life data-parallel applications have complex (non-linear) properties and show that workload distribution has become an important decision variable that can no longer be ignored. Briefly, the total energy consumption during an application execution is the sum of dynamic and static energy consumptions. Static energy consumption is defined as the energy consumed by the platform without the application execution. Dynamic energy consumption is calculated by subtracting this static energy consumption from the total energy consumed by the platform during the application execution. The works \cite{LastovetskyReddy2017,manumachu2018bi,manumachu2018bicpe} propose model-based data partitioning methods that take as input discrete performance and dynamic energy functions with no shape assumptions, which accurately and realistically account for resource contention and NUMA inherent in modern multicore CPU platforms. Using a simulation of the execution of a data-parallel matrix multiplication application based on OpenBLAS DGEMM on a homogeneous cluster of multicore CPUs, it is shown \cite{LastovetskyReddy2017} that optimizing for performance alone results in average and maximum dynamic energy reductions of 24\% and 68\%, but optimizing for dynamic energy alone results in performance degradations of 95\% and 100\%. For a 2D fast Fourier transform application based on FFTW, the average and maximum dynamic energy reductions are 29\% and 55\% and the average and maximum performance degradations are both 100\%. Research work \cite{manumachu2018bi} proposes a solution method to solve bi-objective optimization problem of an application for performance and energy on homogeneous clusters of modern multicore CPUs. This method is shown to determine a diverse set of globally Pareto-optimal solutions whereas existing solution methods give only one solution when the problem size and number of processors are fixed. The methods \cite{LastovetskyReddy2017,manumachu2018bi,manumachu2018bicpe} target homogeneous high performance computing (HPC) platforms. Khaleghzadeh et al. \cite{Hamid2019} propose a solution method solving the bi-objective optimization problem on heterogeneous processors. The authors prove that for an arbitrary number of processors with linear execution time and dynamic energy functions, the globally Pareto-optimal front is linear and contains an infinite number of solutions out of which one solution is load balanced while the rest are load imbalanced. A data partitioning algorithm is presented that takes as an input discrete performance and dynamic energy functions with no shape assumptions. 

\begin{table*}
	\caption{Specifications of the Intel multicore CPUs, HCLServer01-04, ordered by increasing number of sockets and an increasing number of cores per socket.}
	\label{table:hclservers}
	\centering
	\begin{tabular}{ |l|l|l|l|l| }
		\hline
		\textbf{Technical Specifications} & \textbf{HCLServer1 (S1)} & \textbf{HCLServer2 (S2) } & \textbf{HCLServer3 (S3)} & \textbf{HCLServer4 (S4)} \\ \hline
		Processor & Intel Xeon Gold 6152 & Intel Haswell E5-2670V3 & Intel Xeon CPU E5-2699 & Intel Xeon Platinum 8180 \\ \hline
		Core(s) per socket & 22 & 12 & 18 & 28 \\ \hline
		Socket(s) & 1 & 2 & 2 & 2 \\ \hline
		L1d cache, L1i cache  & 32 KB, 32 KB & 32 KB, 32 KB & 32 KB, 32 KB & 32 KB, 32 KB \\ \hline
		L2 cache, L3 cache & 256 KB, 30720 KB & 256 KB, 30976 KB & 256 KB, 46080 KB & 1024 KB, 39424 KB \\ \hline
		Total main memory & 96 GB & 64 GB & 256 GB & 187 GB \\ \hline
		Power meter & WattsUp Pro & WattsUp Pro & - & Yokogawa WT310 \\ \hline
	\end{tabular}
\end{table*}

The research works \cite{LastovetskyReddy2017,manumachu2018bi,manumachu2018bicpe,Hamid2019} are theoretical demonstrating performance and energy improvements based on simulations of clusters of homogeneous and heterogeneous nodes. Khokhriakov et al. \cite{khokhriakov2018performance} present two novel optimization methods to improve the average performance of the FFT routines on modern multicore CPUs. The methods employ workload distribution as the decision variable and are based on parallel computing employing threadgroups. They utilize load imbalancing data partitioning technique that determines optimal workload distributions between the threadgroups, which may not load-balance the application in terms of execution time. The inputs to the methods are discrete 3D functions of performance against problem size of the threadgroups, and can be employed as nodal optimization techniques to construct a 2D FFT routine highly optimized for a dedicated target multicore CPU. The authors employ the methods to demonstrate significant performance improvements over the basic FFTW and Intel MKL FFT 2D routines on a modern Intel Haswell multicore CPU consisting of thirty-six physical cores.

The findings in \cite{LastovetskyReddy2017,manumachu2018bi,manumachu2018bicpe,khokhriakov2018performance,Hamid2019} motivate us to study the influence of three-dimensional decision variable space on bi-objective optimization of applications for performance and energy on multicore CPUs. The three decision variables are: a). The number of identical multithreaded kernels (threadgroups) involved in the parallel execution of an application; b). The number of threads in each threadgroup; and c). The workload distribution between the threadgroups. We focus exclusively on the first two decision variables in this work. The number of possible workload distributions increases exponentially with increasing number of threadgroups employed in the execution of a data-parallel application and it would require employment of threadgroup-specific performance and energy models to reduce the complexity. It is a subject of our future work. 

We propose and study the first application-level method for bi-objective optimization of multithreaded data-parallel applications on a single multicore CPU for performance and energy. The method uses two decision variables, the number of identical multithreaded kernels (threadgroups) executing the application in parallel and the number of threads in each threadgroup. The workload distribution is not a decision variable. It is fixed so that a given workload is always partitioned equally between the threadgroups. The method allows full reuse of highly optimized scientific codes and does not require any changes to hardware or OS. The first step of the method includes writing a data-parallel version of the base kernel that can be executed using a variable number of threadgroups in parallel and solving the same problem as the base kernel, which employs one threadgroup.

We demonstrate our method using four multithreaded applications: a) 2D-FFT using FFTW 3.3.7; b) 2D-FFT using Intel MKL FFT; c) Dense matrix-matrix multiplication using OpenBLAS; and d) Dense matrix-matrix multiplication using Intel MKL FFT. 

Four different modern Intel multicore CPUs are used in the experiments:
a) A single-socket Intel Skylake consisting of 22 physical cores; b) A dual-socket Intel Haswell consisting of 24 physical cores; c) A dual-socket Intel Haswell consisting of 36 physical cores; and d) A dual-socket Intel Skylake consisting of 56 cores. Specifications of the experimental servers S1, S2, S3, and S4 
equipped with these CPUs are given in  Table \ref{table:hclservers}. Servers S1, S2, and S4 are equipped with power meters and fully instrumented for system-level energy measurements. Server S3 is not equipped with a power meter and therefore is not employed in the experiments for single-objective optimization for energy and bi-objective optimization for performance and energy.

\begin{figure}[!t]
\includegraphics[width=1\linewidth]{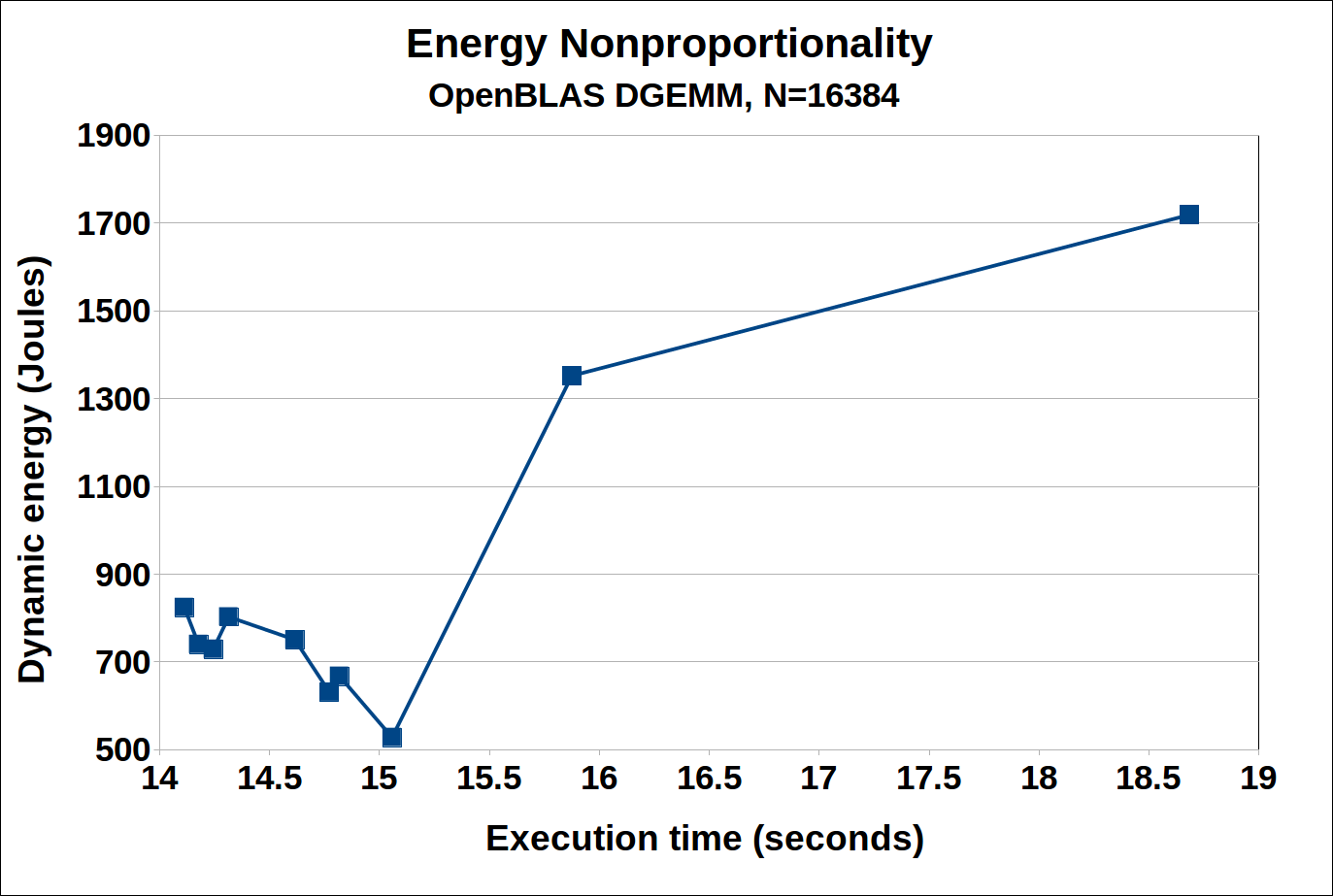}
\caption{Energy nonproportionality on S2 found by our method for OpenBLAS DGEMM application solving workload size, N=16384.}
\label{fig:Energy_nonproportionality_16384}
\end{figure}

Figure \ref{fig:Energy_nonproportionality_16384} illustrates the energy nonproportionality on S2 found by our method for OpenBLAS DGEMM application solving workload size, N=16384. Data points in the graph represent different configurations of the multithreaded application solving exactly the same problem. Energy proportionality is signified by a monotonically increasing relationship between energy and execution time. This is clearly not the case for the relationship shown in the figure.

The average and maximum performance improvements using the number of threadgroups and the number of threads per group as decision variables for performance optimization on a single-socket multicore CPU (S1) are (7\%, 26.3\%), (5\%, 6.5\%) and (27\%, 69\%) for the OpenBLAS DGEMM, Intel MKL DGEMM and Intel MKL FFT applications against their best single threadgroup configurations. Along with performance optimization, the energy improvements for OpenBLAS DGEMM and Intel MKL DGEMM are (7.9\%, 30\%) and (35.7\%, 67\%) against their best single threadgroup configurations.

At the same time, the optimization for performance alone results in average and maximum increases in dynamic energy consumption of (22.5\%, 67\%) and (87\%, 89\%) for the Intel MKL DGEMM and Intel MKL FFT applications in comparison with their energy-optimal configurations. The optimization for dynamic energy alone results in average and maximum performance degradations of (27\%, 39\%) and (19.7\%, 38.2\%) in comparison with their performance-optimal configurations. The average and the maximum number of globally Pareto-optimal solutions for Intel MKL DGEMM and Intel MKL FFT are (2.3, 3) and (2.6, 3).  

On the 24-core dual-socket CPU (S2), the average and maximum performance improvements of (16\%, 20\%) and (8\%, 21\%) for the OpenBLAS DGEMM and Intel MKL DGEMM applications against their best single-threadgroup configurations. Even higher average and maximum performance improvements of (30\%, 50\%) are achieved for the FFTW application on the 56-core dual-socket CPU (S4). Again, the improvements are measured against the original single-threadgroup basic routine employing optimal number of threads. 

At the same time, we find that optimization of the OpenBLAS DGEMM and Intel MKL DGEMM applications on S2 for performance only, results in average and maximum increases in dynamic energy consumption of (15\%, 35\%) and (7.1\%, 49\%) in comparison with their energy-optimal configurations, and optimization of the Intel MKL FFT and FFTW applications on S4 for performance alone results in average and maximum increases in dynamic energy consumption of (7\%, 25\%) and (15\%, 57\%). 

On S2, the optimization of the OpenBLAS DGEMM and Intel MKL DGEMM applications for energy only, results in average and maximum performance degradations of (2.5\%, 6\%) and (3.7\%, 11\%). On S4, the average and maximum performance degradations are (20\%, 33\%) and (31\%, 49\%) for the Intel MKL FFT and FFTW applications. The performance degradations are over the performance-optimal configuration. 

By solving the bi-objective optimization problem on three servers \{S1,S2,S4\}, the average and the maximum number of globally Pareto-optimal solutions determined by out method are (2.7, 3), (3,11), (2.4, 5) and (1.8, 4) for Intel MKL FFT, FFTW, OpenBLAS DGEMM and Intel MKL DGEMM applications. Finally, we propose a qualitative dynamic energy model based on linear regression and employing performance monitoring counters (PMCs) as parameters, which we use to explain the discovered energy nonproportionality and the Pareto-optimal solutions determined by our method. 

The main contributions in this work are the following:

\begin{itemize}
	\item We show that energy proportionality does not hold true for multicore CPUs thereby affording an opportunity for bi-objective optimization for performance and energy. 
	\item We propose and study the first application-level  method for bi-objective optimization of multithreaded data-parallel applications for performance and energy. The method uses two decision variables, the number of identical multithreaded kernels (threadgroups) and the number of threads in each threadgroup. Using four highly optimized data-parallel applications, the proposed method is shown to determine good numbers of globally Pareto-optimal configurations of the applications providing the programmers better trade-offs between performance and energy consumption.
	\item A qualitative dynamic energy model based on linear regression and employing performance monitoring counters (PMCs) as parameters is proposed to explain the Pareto-optimal solutions determined by our solution method for multicore CPUs. The model shows that the energy nonproportionality on our experimental platforms for the two data-parallel applications is due to disproportionately high energy consumption by the data translation lookaside buffer (dTLB) activity.
\end{itemize}

The rest of the paper is organized as follows. Section \ref{sec:related_work} presents the related work. Section \ref{sec:mop} contains brief background on multi-objective optimization and the concept of Pareto-optimality. Section \ref{sec:boppe_solution_method} describes our solution method. Section \ref{sec:boppetg_mm} describes the first step of our solution method for two data-parallel applications, 2D fast Fourier transform and matrix-matrix multiplication. Section \ref{sec:exp_results} contains the experimental results. Section \ref{sec:pmcs} presents our dynamic energy model employing PMCs as parameters to explain the cause behind the energy nonproportionality on our experimental platforms. Section \ref{sec:conclusions} concludes the paper.

\section{Related Work} \label{sec:related_work}

We present an overview of single-objective optimization solution methods for performance or energy followed by bi-objective optimization solution methods for both performance and energy on multicore CPU platforms. Energy models of computing complete the section.

\subsection{Performance Optimization}

There are three dominant approaches in this category. First category contains research works \cite{Fedorova2007,Zhuravlev2010} that have proposed contention-aware thread-level schedulers that try to minimize performance losses due to contention for on-chip shared resources.

The second category includes DRAM controller schedulers that aim to efficiently utilize the shared resource, which is the DRAM memory system, and last level cache partitioning that physically partition the shared resources to minimize contention. DRAM controller schedulers \cite{Ebrahimi2011,Jeong2012} improve the throughput by ordering threads and prioritizing their memory requests through DRAM controllers. Last level cache partitioners \cite{Lin2008,Tam2009} explicitly partition the cache when the default cache replacement policies (such as least-recently-used (LRU)) do not result in efficient execution of applications. These partitioners, however, must be used in conjunction with schedulers that mitigate contention for memory controllers and on-chip interconnects.

The final category includes research works that focus on thread-level schedulers that exploit data sharing between the threads to co-schedule them \cite{Tang2011,Mars2011}. A key building work in the schedulers are performance models based on PMCs that can predict performance loss due to co-scheduling or migrating threads between cores.

\subsection{Energy Optimization}

There are three important categories dealing with energy optimization on multicore CPU platforms. The software category contains research works that propose shared resource partitioners. The two hardware categories concern research works that employ Dynamic Voltage and Frequency Scaling (DVFS) and Dynamic Power Management (DPM) and thermal management. Zhuravlev et al. \cite{Zhuravlev2013} survey the prominent works in all the three categories.

Research works \cite{Wang2011,Chen2013} propose dynamic reconfiguration of private caches and partitioning of shared caches (last level cache, for example) to reduce the energy consumption without hurting performance.

DVFS and DPM allow changing the frequencies of the cores and to lower their power states when they are idle. Considering the enormity of literature in this category, we will cover only works that take into account resource contention and thread-to-core mapping while employing DVFS. Kadayif et al. \cite{Kadayif2004} exploit the heterogeneous nature of workloads executed by different processors to set their frequencies so as to reduce energy without impacting performance. Research works \cite{Kondo2007,Watanabe2007} employ DVFS to reduce resource contention and energy consumption. 

The main goal of thermal management algorithms is to find thread-to-core mappings (or even thread migration) to remove drastic variations in temperatures or thermal hotspots in the chip and at the same time reduce the energy consumption without impacting the performance. They employ as inputs thermal models that are built using temperature measurements provided by on-chip sensors \cite{Yang2008,Ayoub2009}. The algorithms are chiefly employed at the OS level.

Asymmetry-aware schedulers have been proposed for energy optimization on asymmetric multicore systems, which feature a mix of fast and slow cores, high-power and low-power cores but that expose the same instruction-set architecture (ISA). Fedorova et al. \cite{Fedorova2009} propose a system-level scheduler that assigns sequential phases of an application to fast cores and parallel phases to slow cores to maximize the energy efficiency. Herbert et al. \cite{Herbert2012} employ DVFS to exploit the core-to-core variations from fabrication in power and performance to improve the energy efficiency of the multicore platform.

\subsection{Optimization for Performance and Energy}

Das et al. \cite{Das2014} propose task mapping to optimize for energy and reliability on multiprocessor systems-on-chip (MPSoCs) with performance as a constraint. Sheikh et al. \cite{Sheikh2016} propose task scheduler employing evolutionary algorithms to optimize applications on multicore CPU platforms for performance, energy, and temperature. Abdi et al. \cite{Abdi2019} propose multi-criteria optimization where they minimize the execution time under three constraints, the reliability, the power consumption, and the peak temperature. DVFS is a key decision variable in all of these research works.

The following research works focus on application-level solution methods. Subramaniam et al. \cite{Subramaniam2010} use multi-variable regression to study the performance-energy trade-offs of the high-performance LINPACK (HPL) benchmark. They study performance-energy trade-offs using the decision variables, number of threads and number of processes. Marszalkowski et al. \cite{Marszałkowski2016} analyze the impact of memory hierarchies on time-energy trade-off in parallel computations, which are represented as divisible loads. They represent execution time and energy by two linear functions on problem size, one for in-core computations and the other for out-of-core computations. 

Research works \cite{LastovetskyReddy2017,manumachu2018bicpe} propose data partitioning algorithms that solve single-objective optimization problems of data-parallel applications for performance or energy on homogeneous clusters of multicore CPUs. They take as an input, discrete performance and dynamic energy functions with no shape assumptions and that accurately and realistically account for resource contention and NUMA inherent in modern multicore CPU platforms. Research work \cite{manumachu2018bi} proposes a solution method to solve bi-objective optimization problem of an application for performance and energy on homogeneous clusters of modern multicore CPUs. They demonstrate that the method gives a diverse set of globally Pareto-optimal solutions and that it can be combined with DVFS-based multi-objective optimization methods to give a better set of (globally Pareto-optimal) solutions. The methods target homogeneous HPC platforms. Chakraborti et al. \cite{chakrabarti2017pareto} consider the effect of heterogeneous workload distribution on bi-objective optimization of data analytics applications by simulating heterogeneity on homogeneous clusters. The performance is represented by a linear function of problem size and the total energy is predicted using historical data tables. Khaleghzadeh et al. \cite{Hamid2019} propose a solution method solving the bi-objective optimization problem on heterogeneous processors and comprising of two principal components. The first component is a data partitioning algorithm that takes as an input discrete performance and dynamic energy functions with no shape assumptions. The second component is a novel methodology employed to build the discrete dynamic energy profiles of individual computing devices, which are input to the algorithm.

\subsection{Energy Predictive Models of Computation}

Energy predictive models predominantly employ performance monitoring counters (PMCs) as parameters. Bellosa et al. \cite{Bellosa2000} propose an energy model based on performance monitoring counters such as integer operations, floating-point operations, memory requests due to cache misses, etc. that they believed to strongly correlate with energy consumption. A linear model that is based on the utilization of CPU, disk, and network is presented in \cite{Heath2005}. A more complex power model (Mantis) \cite{Economou2006} employs utilization metrics of CPU, disk, and network components and hardware performance counters for memory as predictor variables. 

Fan et al. \cite{Fan2007} propose a simple linear model that correlates power consumption of a single-core processor with its utilization. Bertran et al. \cite{Bertran2010} present a power model that provides per-component power breakdown of a multicore CPU. Dargie et al. \cite{Dargie2015} use the statistics of CPU utilization (instead of PMCs) to model the relationship between the power consumption of multicore processor and workload quantitatively. They demonstrate that the relationship is quadratic for single-core processor and linear for multicore processors. Lastovetsky et al. \cite{LastovetskyReddy2017} present an application-level energy model where the dynamic energy consumption of a processor is represented by a discrete function of problem size, which is shown to be highly non-linear for data-parallel applications on modern multicore CPUs.

\section{Multi-Objective Optimization: Background} \label{sec:mop}

A multi-objective optimization (MOP) problem may be defined as follows \cite{Miettinen1999},\cite{Talbi2009}:
\begin{alignat*}{3}
& minimize \quad \{\mathcal{F}(x) = (f_1(x),...,f_k(x))\} & \\
& \text{Subject to} \quad x \in \mathcal{S}
\end{alignat*}
where there are $k (\ge 2)$ objective functions $f_i:\mathbb{R}^p \rightarrow \mathbb{R}$. The objective is to minimize all the objective functions simultaneously.

$\mathcal{F}(x) = (f_1(x),...,f_k(x))^T$ denotes the vector of objective functions. The decision (variable) vectors $x = (x_1,...,x_p)$ belong to the (non-empty) feasible region (set) $\mathcal{S}$, which is a subset of the decision variable space $\mathbb{R}^p$. We call the image of the feasible region represented by $\mathcal{Z}$ ($= f(\mathcal{S})$), the feasible objective region. It is a subset of the objective space $\mathbb{R}^k$. The elements of $\mathcal{Z}$ are called objective (function) vectors or criterion vectors and denoted by $\mathcal{F}(x)$ or $z = (z_1,...,z_k)^T$, where $z_i = f_i(x), \forall i \in [1,k]$ are objective (function) values or criterion values.

If there is no conflict between the objective functions, then a solution $x^*$ can be found where every objective function attains its optimum \cite{Talbi2009}.
\begin{alignat*}{3}
\forall x \in \mathcal{S}, f_i(x^*) \leq f_i(x), \quad i = 1,...,k
\end{alignat*}
However, in real-life multi-objective optimization problems, the objective functions are at least partly conflicting. Because of this conflicting nature of objective functions, it is not possible to find a single solution that would be optimal for all the objectives simultaneously. In multi-objective optimization, there is no natural ordering in the objective space because it is only partially ordered. Therefore we must treat the concept of optimality differently from single-objective optimization problem. The generally used concept is \textit{Pareto-optimality}.

\begin{definition}
	A decision vector $x^* \in \mathcal{S}$ is \textit{Pareto-optimal} if there does not exist another decision vector $x \in \mathcal{S}$ such that $f_i(x) \leq f_i(x^*), \forall i = 1,...,k$ and $f_j(x) < f_j(x^*)$ for at least one index $j$ \cite{Miettinen1999}.
\end{definition}

An objective vector $z^* \in \mathcal{Z}$ is Pareto-optimal if there does not exist another objective vector $z \in \mathcal{Z}$ such that $z_i \leq z_i^*, \forall i = 1,...,k$ and $z_j < z_j^*$ for at least one index $j$.

\begin{definition}
	A decision vector $x^* \in \mathcal{S}$ is \textit{weakly Pareto-optimal} if there does not exist another decision vector $x \in \mathcal{S}$ such that $f_i(x) < f_i(x^*), \forall i = 1,...,k$ \cite{Miettinen1999}.
\end{definition}

An objective vector $z^* \in \mathcal{Z}$ is Pareto-optimal if there does not exist any other vector for which all the component objective vector values are better.

Mathematically speaking, every Pareto-optimal point is an equally acceptable solution of the multi-objective optimization problem. Therefore, user preference relations (or preferences of decision maker) are provided as input to the solution process to select one or more points from the set of Pareto-optimal solutions \cite{Miettinen1999}.

\begin{figure}[!t]
	\centering
	\includegraphics[width=3.5in]{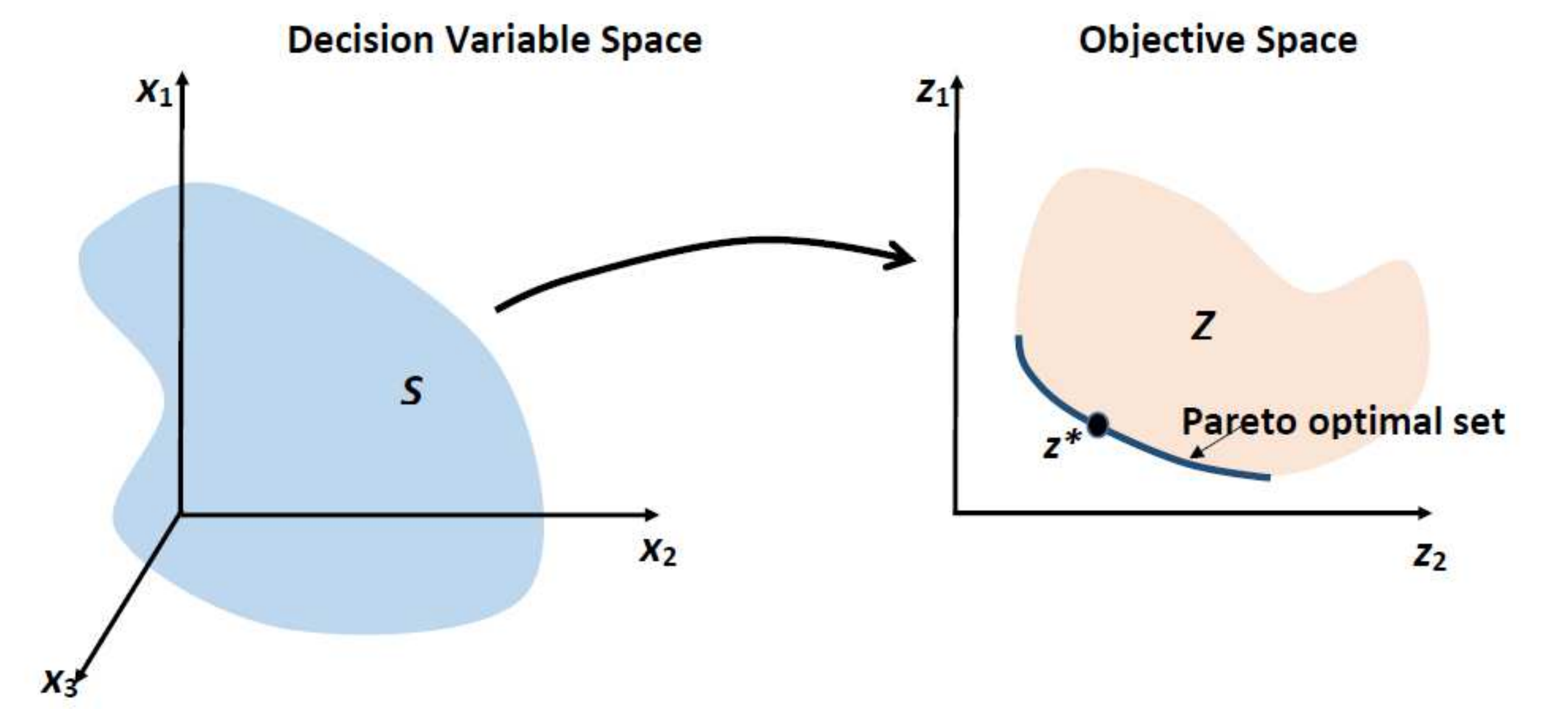}
	\caption{An example showing the set $\mathcal{S}$ of decision variable vectors, the set $\mathcal{Z}$ of objective vectors, and Pareto-optimal objective vectors shown by bold line. $\mathcal{S} \subset \mathbb{R}^3, \mathcal{Z} \subset \mathbb{R}^2$.}
	\label{fig:paretooptimality}
\end{figure}

\begin{figure*}[!t]
	\centering
	\subfloat[][]{
		\includegraphics[width=0.5\linewidth]{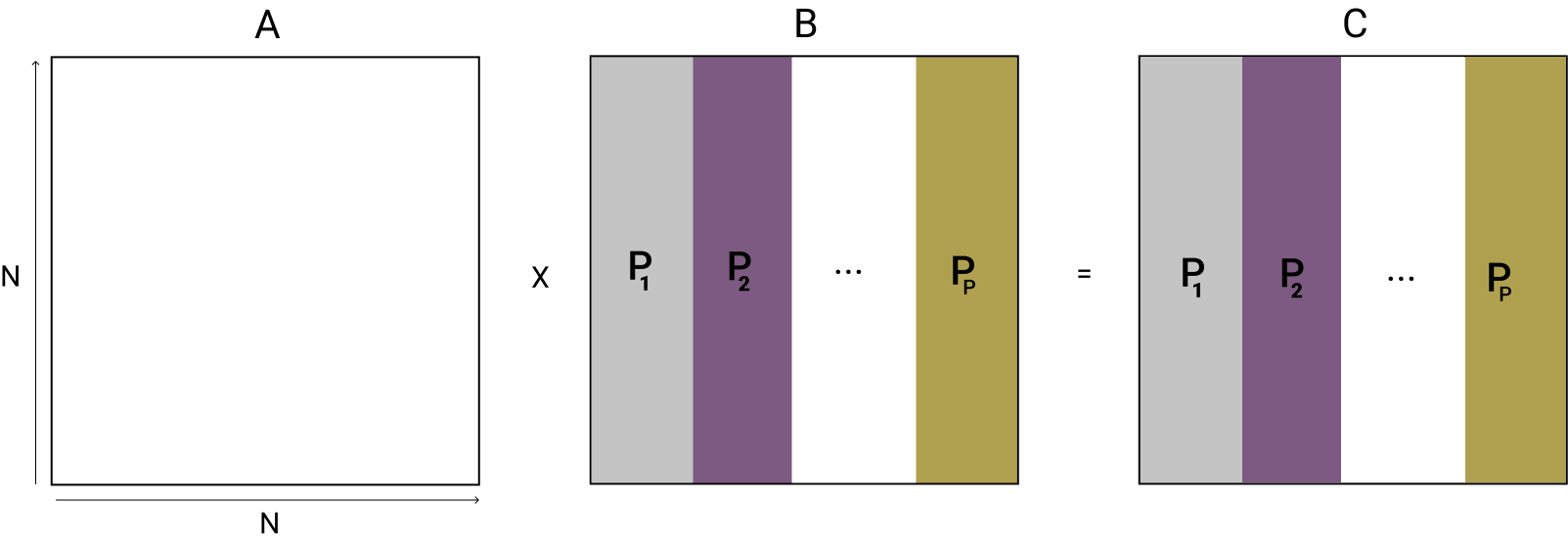}
		\label{fig:pmm-group-lb-v}}
	\hfill
	\subfloat[][]{
		\includegraphics[width=0.5\linewidth]{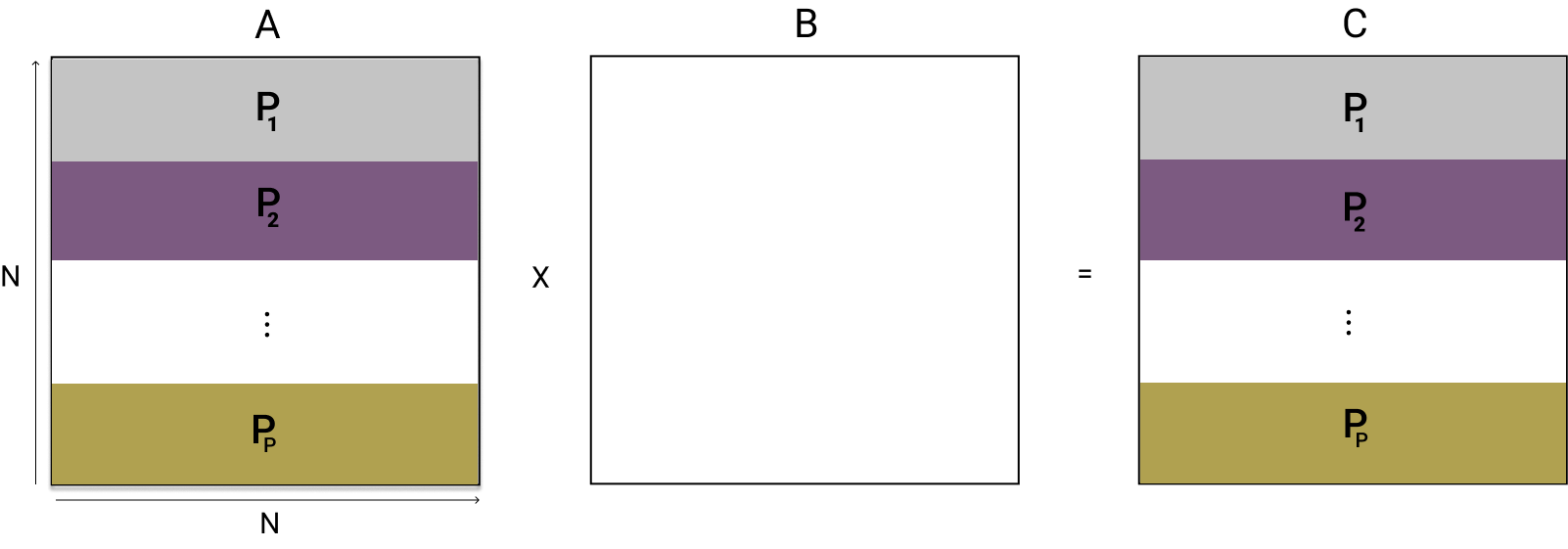}
		\label{fig:pmm-group-lb-h}}
	\hfill
	\subfloat[][]{
		\includegraphics[width=0.5\linewidth]{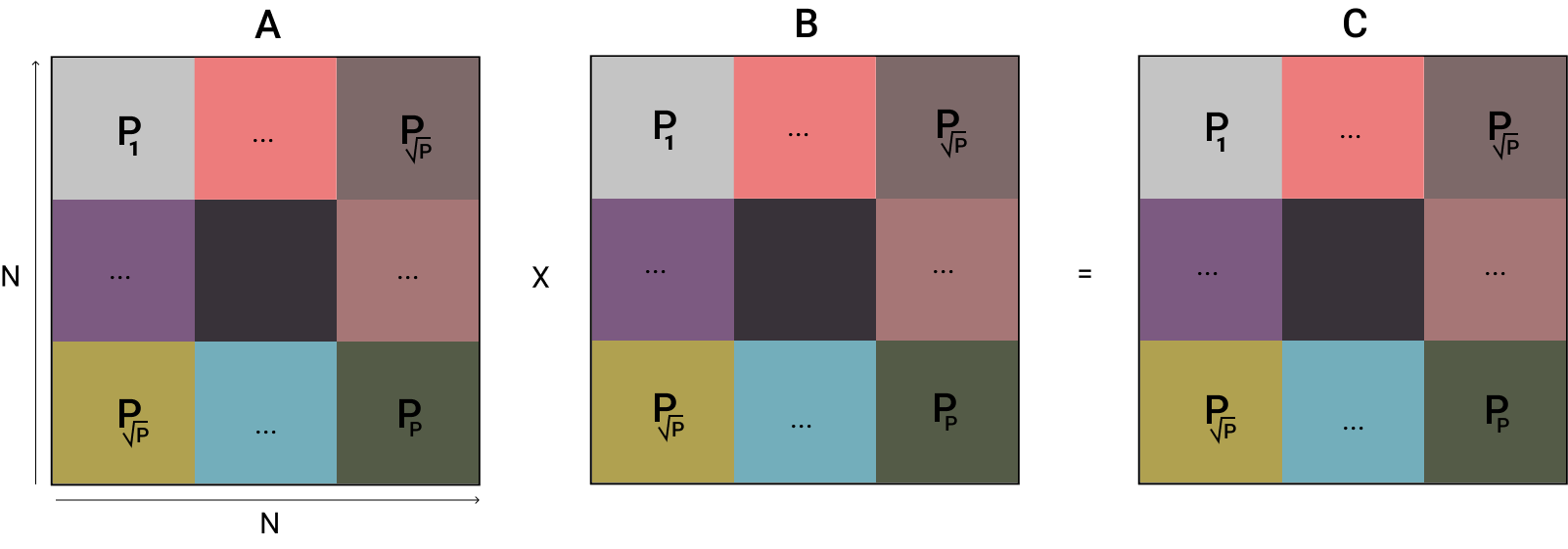}
		\label{fig:pmm-group-lb-s}}
	\hfill
	\caption{(a). PMMTG-V: Matrices B and C are vertically partitioned among the threadgroups. (b). PMMTG-H: Matrices A and C are horizontally partitioned among the threadgroups. (c). PMMTG-S: The $p$ threadgroups are arranged in a square grid of size $\sqrt p \times \sqrt p$. All the matrices are partitioned into squares among the threadgroups.}
\end{figure*}

\begin{figure*}[!t]
	\centering
	\includegraphics[width=0.7\linewidth]{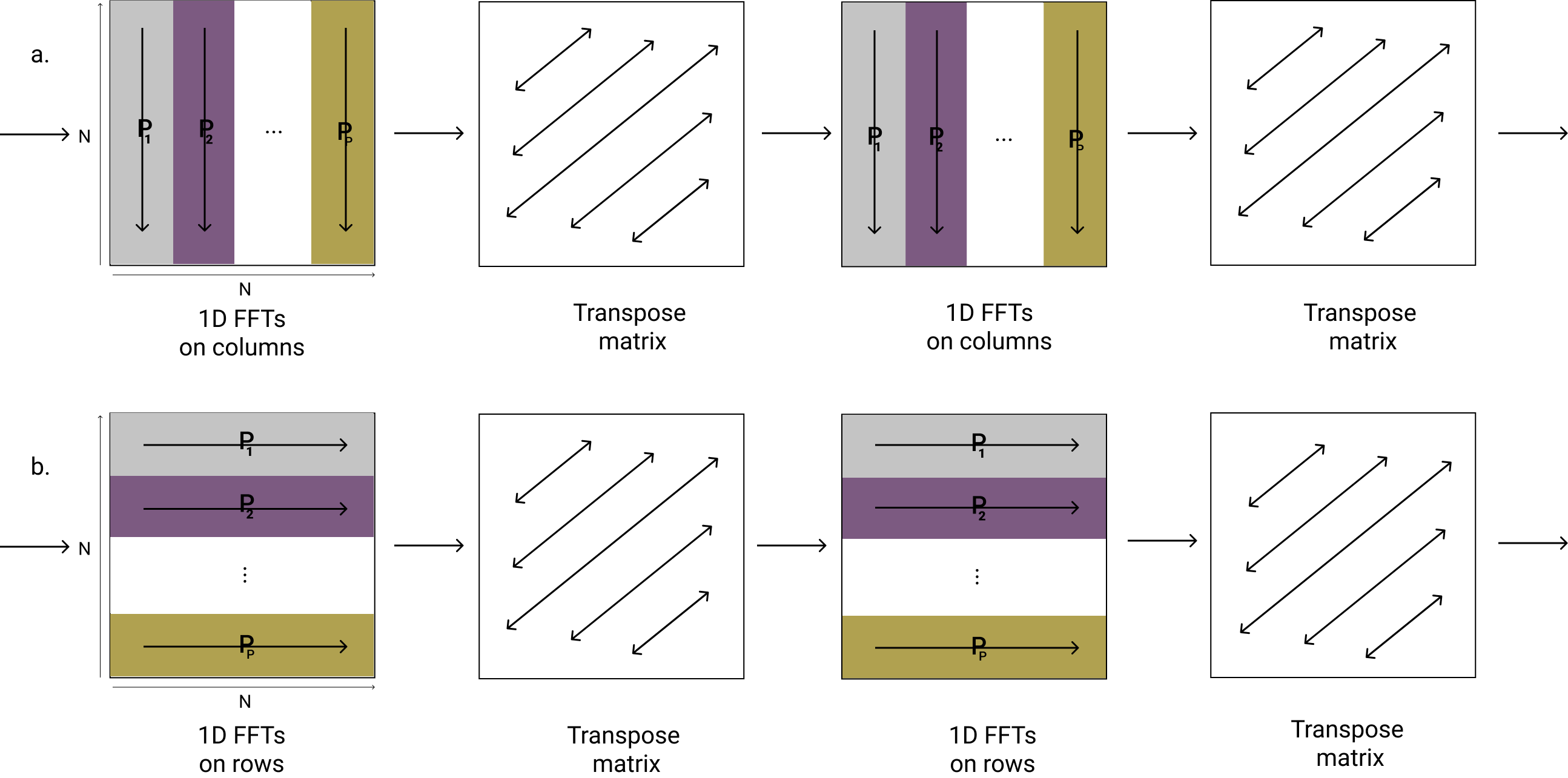}
	\caption{2D-DFT of signal matrix M of size $N \times N$ using $p$ threadgroups. a). PFFTTG-V using vertical decomposition of the signal matrix. b). PFFTTG-H using horizontal decomposition of the signal matrix.}
	\label{fig:pfft-group-lb-h}
\end{figure*}

In Figure \ref{fig:paretooptimality}, a feasible region $\mathcal{S} \subset \mathbb{R}^3$ and its image, a feasible objective region $\mathcal{Z} \subset \mathbb{R}^2$, are shown. The thick blue line in the figure showing the objective space contains all the Pareto-optimal objective vectors. The vector $z^*$ is one of them.

In this work, we consider bi-objective optimization where performance and dynamic energy are the objectives.

\section{Solution Method Solving Bi-objective Optimization Problem on a Single Multicore CPU} \label{sec:boppe_solution_method}

In this section, we describe our solution method, BOPPETG, for solving the bi-objective optimization problem of a multithreaded data-parallel application on multicore CPUs for performance and energy (BOPPE). The method uses two decision variables, the number of identical multithreaded kernels (threadgroups) and the number of threads in each threadgroup. A given workload is always partitioned equally between the threadgroups.

The bi-objective optimization problem (BOPPE) can be formulated as follows: Given a multithreaded data-parallel application of workload size $n$ and a multicore CPU of $l$ cores, the problem is to find a globally Pareto-optimal front of solutions optimizing execution time and dynamic energy consumption during the parallel execution of the workload. Each solution is an application configuration given by (threadgroups, threads per group).

The inputs to the solution method are the workload size of the multi-threaded data-parallel application, $n$; the number of cores in the multicore CPU, $l$; the multithreaded base kernel, $mtkernel$; the base power of the multicore CPU platform, $P_b$. The outputs are the globally Pareto-optimal front of objective solutions, $\mathcal{P}_{opt}$, and the optimal application configurations corresponding to these solutions, $\mathcal{C}_{opt}$. Each Pareto-optimal solution of objectives $o$ is represented by the pair, $(s_o,e_o)$, where $s_o$ is the execution time and $e_o$ is the dynamic energy. Associated with this solution is an array of application configurations, $\mathcal{A}(g_o,t_o)$, containing decision variable pairs, $(g_o,t_o)$, where $g_o$ represents the number of threadgroups each containing $t_o$ threads.

The main steps of BOPPETG are as follows:

\textbf{Step 1. Parallel implementation allowing ($g$,$t$) configuration:} Design and implement a data-parallel version of the base kernel $mtkernel$ and that can be executed using $g$ identical multithreaded kernels in parallel. Each kernel is executed by a threadgroup containing $t$ threads. The workload $n$ is divided equally between the $g$ threadgroups during the execution of the data-parallel version. The data-parallel version should essentially allow its runtime configuration using number of threadgroups and number of threads per group with the workload equally partitioned between the threadgroups. 

\textbf{Step 2. Initialize $g$ and $t$:} All the runtime configurations, ($g$,$t$), where the product, $g \times t$, is less than or equal to the total number of cores ($l$) in the multicore platform are considered. $g \gets 1$, $t \gets 1$. Go to Step 3.

\textbf{Step 3. Determine time and dynamic energy of the ($g$,$t$) configuration of the application:} The data-parallel version composed in Step 1 is run using the ($g$,$t$) configuration. Its execution time and dynamic energy consumption are determined as follows: $s_o = t_f - t_i$, $e_o = e_f - P_b \times s_o$, where $t_i$ and $t_f$ are the starting and ending execution times and $e_f$ is the total energy consumption during the execution of the application. Go to Step 4.

\textbf{Step 4. Update Pareto-optimal front for ($g$,$t$):} The solution $(s_o,e_o)$ if Pareto-optimal is added to the globally Pareto-optimal set of objective solutions, $\{\mathcal{P}_{opt}\}$, and existing member solutions of the set that are inferior to it are removed. The optimal application configurations corresponding to the solution $(s_o,e_o)$ are stored in $\mathcal{C}_{opt}$. Go to Step 5.

\textbf{Step 5. Test and Increment ($g$,$t$):} If $t < l$, $t \gets t + 1$, go to Step 3. Set $g \gets g + 1$, $t \gets 1$. If $g \times t \le l$, go to Step 3. Else return the globally Pareto-optimal front and optimal application configurations given by $\{\mathcal{P}_{opt},\mathcal{C}_{opt}\}$ and quit.

In the following section, we illustrate the first step of BOPPETG for two applications, matrix-matrix multiplication and 2D fast Fourier transform. We show in particular how BOPPETG can reuse highly optimized scientific kernels with careful design and development of parallel versions of the application.

\section{Parallel Matrix-Matrix Multiplication} \label{sec:boppetg_mm}

We illustrate the first step of our solution method (BOPPETG) for implementing the data-parallel version of dense matrix-matrix multiplication (PMMTG).

\begin{figure}[!t]
\lstset{language=C++}
\begin{lstlisting}
void *dgemm(void *input)
{
	int i = *(int*)input;
	openblas_set_num_threads(t);
	goto_set_num_threads(t);
	omp_set_num_threads(t);
	if (i == 1)
	{
       cblas_dgemm(CblasRowMajor, CblasNoTrans,
            CblasNoTrans, N/p, N, N, alpha, A1, N, 
            B, N, beta, C1, N);
	}
    ...
	if (i == p)
	{
       cblas_dgemm(CblasRowMajor, CblasNoTrans, 
            CblasNoTrans, N/p, N, N, alpha, Ap, N, 
            B, N, beta, Cp, N);
	}
}

int main() {
	int row;
#pragma omp parallel for num_threads(p*t)
	for (row = 0; row < N/p; row++) {
		memcpy(&A1[row*N], &A[row*N], N*sizeof(double));
		...
		memcpy(&Ap[row*N], &A[(p-1)*N*(N/p)+row*N], 
		       N*sizeof(double));
		memcpy(&C1[row*N], &C[row*N], N*sizeof(double));
		...
		memcpy(&Cp[row*N], &C[(p-1)*N*(N/p)+row*N], 
		       N*sizeof(double));
	}
	
	pthread_t t1, ..., tp;
	int i1 = 1, ..., ip = p;
	pthread_create(&t1, NULL, dgemm, &i1);
	...
	pthread_create(&tp, NULL, dgemm, &ip);
	pthread_join(tp, NULL);
	...
	pthread_join(t1, NULL);
	
#pragma omp parallel for num_threads(p*t)
	for (row = 0; row < N/p; row++)
	{
		memcpy(&A[row*N], &A1[row*N], N*sizeof(double));
		...
		memcpy(&A[(p-1)*N*(N/p)+row*N], &Ap[row*N], 
		       N*sizeof(double));
		memcpy(&C[row*N], &C1[row*N], N*sizeof(double));
		...
		memcpy(&C[(p-1)*N*(N/p)+row*N], &Cp[row*N], 
		       N*sizeof(double));
	}
}
\end{lstlisting}
\caption{OpenBLAS implementation of parallel matrix-matrix multiplication using horizontal decomposition (PMMTG-H) and employing $p$ threadgroups of $t$ threads each.}
\label{fig:pmm-group-lb-h-code}
\end{figure}

The PMMTG application computes the matrix product ($C = \alpha \times A \times B + \beta \times C$) of two dense square matrices A and B of size $N \times N$. The application is executed using $p$ threadgroups, $\{P_1,...,P_p\}$. To simplify the exposition of the algorithms, we assume N to be divisible by p. 

There are three parallel algorithmic variants of PMMTG. In PMMTG-V, the matrices B and C are partitioned vertically such that each threadgroup is assigned $\frac{N}{p}$ of the columns of B and C as shown in the Figure \ref{fig:pmm-group-lb-v}. Each threadgroup $P_i$ computes its vertical partition $C_{P_i}$ using the matrix product, $C_{P_i} = \alpha \times A \times B_{P_i} + \beta \times C_{P_i}$. In PMMTG-H, the matrices A and C are partitioned horizontally such that each threadgroup is assigned $\frac{N}{p}$ of the rows of B and C as shown in the Figure \ref{fig:pmm-group-lb-h}. Each threadgroup $P_i$ computes its horizontal partition $C_{P_i}$ using the matrix product, $C_{P_i} = \alpha \times A_{P_i} \times B + \beta \times C_{P_i}$. In PMMTG-S, the $p$ threadgroups $\{P_1,...,P_p\}$ are arranged in a square grid $Q_{st}, s \in [1,\sqrt p], t \in [1,\sqrt p]$. The matrices A, B, and C are partitioned into equal squares among the threadgroups as shown in the Figure \ref{fig:pmm-group-lb-s}. In each matrix, each threadgroup $P_i (= Q_{st})$ is assigned a sub-matrix of size $\frac{N}{\sqrt p} \times \frac{N}{\sqrt p}$ and computes its square partition $C_{Q_{st}}$ using the matrix product, $C_{Q_{st}} = \alpha \times \sum_{k=1}^{\sqrt{p}}(A_{sk} \times B_{kt}) + \beta \times C_{Q_{st}}$. $A_{sk}$ is the square block in matrix $A$ located at $(s,k)$. $B_{kt}$ is the square block in matrix $B$ located at $(k,t)$.

\subsection{Implementation of PMMTG-H Based on OpenBLAS DGEMM}

We describe an OpenBLAS implementation of PMMTG-H (Figure \ref{fig:pmm-group-lb-h-code}) here. The implementations of the other PMMTG algorithms employing Intel MKL and OpenBLAS are described in the supplemental.

The inputs to an implementation are: a). Matrices A, B, and C of sizes $N \times N$; b). Constants $\alpha$ and $\beta$; c) The number of threadgroups, $\{P_1,\cdots,P_p\}$; d). The number of threads in each threadgroup represented by $t$. The output matrix, C, contains the matrix product.

The vertical partitions of A and C, \{$A_{P_i},C_{P_i}$\}, $i \in [1,p]$, assigned to the threadgroups, $\{P_1,...,P_p\}$, are initialized in Lines 24-34. Then $p$ pthreads representing the $p$ threadgroups are created, each a multithreaded OpenBLAS DGEMM kernel executing $t$ OpenMP threads (Lines 36-43).The $p$ threadgroups compute the matrix-matrix product (Lines 1-20). The result is gathered in the matrix C (Lines 45-56).

The implementations using Intel MKL differ from those using OpenBLAS. In Intel MKL, the matrix-matrix computation by a threadgroup is performed using an OpenMP parallel region with $t$ threads whereas the same is done in OpenBLAS using a pthread.

\section{Parallel 2D Fast Fourier Transform} \label{sec:boppetg_fft}

We present here the first step of our solution method (BOPPETG) to compose the data-parallel version of 2D Fast Fourier Transform (PFFTTG). The sequential 2D FFT algorithm is described first before the two parallel algorithmic variants of 2D Fast Fourier Transform.

The definition of 2D-DFT of a two-dimensional point discrete signal M of size $N \times N$ is below:
\begin{align*}
M[k][l] &= \sum_{i=0}^{N-1}\sum_{j=0}^{N-1} M[i][j] \times \omega_N^{ki} \times \omega_N^{lj} \\
\omega_N &= e^{-\frac{2\pi}{N}}, 0 \le k,l \le N-1
\end{align*}

M is the signal matrix where each element $M[i][j]$ is a complex number. The total number of complex multiplications required to compute the 2D-DFT is $\Theta(N^4)$.

The sequential \textit{row-column decomposition method} reduces this complexity by computing the 2D-DFT using a series of 1D-DFTs, which are implemented using a fast 1D-FFT algorithm. The method consists of two phases called the row-transform phase and column-transform phase. Figure 4 depicts the method, which is mathematically summarized below:

\begin{align*}
M[k][l] &= \sum_{i=0}^{N-1}\sum_{j=0}^{N-1} M[i][j] \times \omega_N^{ki} \times \omega_N^{lj} \\
&= \sum_{i=0}^{N-1} \omega_N^{ki} \times (\sum_{j=0}^{N-1} M[i][j] \times \omega_N^{lj}) \\
&= \sum_{i=0}^{N-1} \omega_N^{ki} \times (\tilde{M}[i][l]) \\
&= \sum_{i=0}^{N-1} (\tilde{M}[i][l]) \times \omega_N^{ki} \\
\omega_N &= e^{-\frac{2\pi}{N}}, 0 \le k,l \le N-1
\end{align*}

It computes a series of ordered 1D-FFTs of size N on the N rows. That is, each row i (of length N) is transformed via a fast 1D-FFT to $\tilde{M}[i][l], \forall l \in [0,N-1]$. The total cost of this row-transform phase is $\Theta(N^2\log_2N)$. Then, it computes a series of ordered 1D-FFTs on the N columns of $\tilde{M}$. The column $l$ of $\tilde{M}$ is transformed to $M[k][l], \forall k \in [0,N-1]$. The total cost of this column-transform phase is $\Theta(N^2\log_2N)$. Thus, by using the \textit{row-column decomposition method}, the complexity of 2D-FFT is reduced from $\Theta(N^4)$ to $\Theta(N^2\log_2N)$. All the FFTs that we discuss in this work are considered in-place.

The PFFTTG application employing our solution method computes the 2D-DFT of the signal matrix of size $N \times N$ using $p$ threadgroups, $\{P_1,...,P_p\}$. It is based on the sequential 2D-FFT row-column decomposition method. There are two parallel algorithmic variants of PFFTTG, PFFTTG-H and PFFTTG-V. To simplify the exposition of the algorithms, we assume N to be divisible by p.

\subsection{PFFTTG-H: Using Horizontal Decomposition of Signal Matrix $M$}

The parallel 2D-FFT algorithm, PFFTTG-H, consists of four steps:

\textbf{Step 1. 1D-FFTs on rows:} Threadgroup $P_i$ executes sequential 1D-FFTs on rows $(i-1) \times \frac{N}{p}+1,...,i \times \frac{N}{p}$.
 
\textbf{Step 2. Matrix Transposition:} Transpose the matrix M.

\textbf{Step 3. 1D-FFTs on rows:} Threadgroup $P_i$ executes sequential 1D-FFTs on rows $(i-1) \times \frac{N}{p}+1,...,i \times \frac{N}{p}$.

\textbf{Step 4. Matrix Transposition:} Transpose the matrix M.

The computational complexity of Steps 1 and 3 is $\Theta(\frac{N^2}{p}\log_2N)$. The computational complexity of Steps 2 and 4 is 
$\Theta(\frac{N^2}{p})$. Therefore, the total computational complexity of PFFTTG-H is $\Theta(\frac{N^2}{p}\log_2N)$.

The algorithm is illustrated in the Figure \ref{fig:pfft-group-lb-h}.

\subsection{PFFTTG-V: Using Vertical Decomposition of Signal Matrix $M$}

The parallel 2D-FFT algorithm, PFFTTG-V, consists of four steps:

\textbf{Step 1. 1D-FFTs on columns:} Threadgroup $P_i$ executes sequential 1D-FFTs on columns $(i-1) \times \frac{N}{p}+1,...,i \times \frac{N}{p}$.
 
\textbf{Step 2. Matrix Transposition:} Transpose the matrix M.

\textbf{Step 3. 1D-FFTs on columns:} Threadgroup $P_i$ executes sequential 1D-FFTs on columns $(i-1) \times \frac{N}{p}+1,...,i \times \frac{N}{p}$.

\textbf{Step 4. Matrix Transposition:} Transpose the matrix M.

The computational complexity of Steps 1 and 3 is $\Theta(\frac{N^2}{p}\log_2N)$. The computational complexity of Steps 2 and 4 is $\Theta(\frac{N^2}{p})$. Therefore, the total computational complexity of PFFTTG-V is $\Theta(\frac{N^2}{p}\log_2N)$.

The algorithm is illustrated in the Figure \ref{fig:pfft-group-lb-h}.

\begin{figure}[!t]
\lstset{language=C++}
\begin{lstlisting}
fftw_plan fftw1d_init_plan(const int sign, const int m,
    const int n, fftw_complex* X, fftw_complex* Y)
{
    int rank = 1, howmany = m;
    int s[] = {n}, idist = n;
    int odist = n, istride = 1;
    int ostride = 1, *inembed = s, *onembed = s;
    return fftw_plan_many_dft(rank, s, howmany,
               X, inembed, istride, idist, Y, onembed, 
               ostride, odist, sign, FFTW_ESTIMATE);
}
int fftw2d(const int sign, const int p, const int N,
    const unsigned int t, const unsigned int blockSize,
    fftw_complex* X
)
{
    fftw_init_threads();
    fftw_plan_with_nthreads(t);
    fftw_plan plan1, plan2, ..., planp;
    plan1 = fftw1d_init_plan(sign, N/p, N, X, X);
    plan2 = fftw1d_init_plan(sign, N/p, N, 
               &X[(N/p)*N], &X[(N/p)*N]);
    ...
    planp = fftw1d_init_plan(sign, N-(p-1)*(N/p), N, 
               &X[(p-1)*(N/p)*N], &X[(p-1)*(N/p)*N]);
#pragma omp parallel sections num_threads(p)
{
    #pragma omp section
    {
       fftw_execute(plan1);
       fftw_destroy_plan(plan1);
    }
...
    #pragma omp section
    {
       fftw_execute(plan12);
       fftw_destroy_plan(plan12);
    }
}
    hcl_transpose_block(X, 0, N, N, t, blockSize);
    plan1 = fftw1d_init_plan(sign, N/p, N, X, X);
    plan2 = fftw1d_init_plan(sign, N/p, N, 
               &X[(N/p)*N], &X[(N/p)*N]);
    ...
    planp = fftw1d_init_plan(sign, N-(p-1)*(N/p), N, 
               &X[(p-1)*(N/p)*N], &X[(p-1)*(N/p)*N]);
#pragma omp parallel sections num_threads(p)
{
    #pragma omp section
    {
       fftw_execute(plan1);
       fftw_destroy_plan(plan1);
    }
...
    #pragma omp section
    {
       fftw_execute(plan12);
       fftw_destroy_plan(plan12);
    }
}
    hcl_transpose_block(X, 0, N, N, nt, blockSize);
    fftw_cleanup_threads();
}

\end{lstlisting}
\caption{FFTW implementation using horizontal decomposition of signal matrix and executed by $p$ threadgroups of $t$ threads each.}
\label{fig:pfft-group-lb-h-code}
\end{figure}

\subsection{Implementation of PFFTTG-H Based on FFTW}

Figure \ref{fig:pfft-group-lb-h-code} illustrates the FFTW implementation of PFFTTG-H. The shared memory implementations of other PFFTTG algorithms based on Intel MKL and FFTW are described in the supplemental.

The inputs to an implementation are: a). Signal matrix M of size $N \times N$; b). The number of threadgroups, $p$, $\{P_1,\cdots,P_p\}$; c). The number of threads in each threadgroup represented by $t$. The output is the transformed signal matrix M (considering that we are performing in-place FFT).

Lines 17-18 show the initialization of FFTW multithreaded runtime. Lines 19-25 show the creation of $p$ FFT plans, each plan executed by a threadgroup of $t$ threads. Lines 1-11 illustrate the creation of a plan using fftw\_dft\_plan\_many routine. Lines 26-39 show the execution and destruction of the plans (1D-FFTs on rows) by the threadgroups. This is followed by transpose of the signal matrix (Line 40). Lines 41-46 contain the creation of $p$ FFT plans (1D-FFTs on rows) followed by their execution by the threadgroups. Finally, the signal matrix is transposed again (Line 61). The FFTW runtime is then destroyed (Line 62).

The implementations based on Intel MKL differ from those employing FFTW. In FFTW, only plan execution (fftw\_plan\_many\_dft) and plan destruction (fftw\_destroy\_plan) are thread-safe and can be called in an OpenMP parallel region.

\section{Experimental Results and Discussion} \label{sec:exp_results}

In this section, we present our experimental results for matrix-matrix multiplication (PMMTG) and 2D fast Fourier transform (PFFTTG) employing our solution method.

To make sure the experimental results are reliable, we follow a statistical methodology described in the supplemental. Briefly, for every data point in the functions, the automation software executes the application repeatedly until the sample mean lies in the 95\% confidence interval and a precision of 0.025 (2.5\%) has been achieved. For this purpose, Student's t-test is used assuming that the individual observations are independent and their population follows the normal distribution. The validity of these assumptions is verified by plotting the distributions of observations and using Pearson's Test. The speed/time/energy values shown in the graphical plots are the sample means.

Four multicore CPUs shown in the Table \ref{table:hclservers} and described earlier are used in the experiments. Three platforms \{S1, S2, S4\} have a power meter installed between their input power sockets and the wall A/C outlets. S1 and S2 are connected with a \emph{Watts Up Pro} power meter; S4 is connected with a \emph{Yokogawa WT310} power meter. S3 is not equipped with a power meter and therefore is not employed in the experiments for single-objective optimization for energy and bi-objective optimization for performance and energy. 

The power meter provides the total power consumption of the server. It has a data cable connected to one USB port of the server. A script written in Perl collects the data from the power meter using the serial USB interface. The execution of the script is non-intrusive and consumes insignificant power. WattsUp Pro power meters are periodically calibrated using the ANSI C12.20 revenue-grade power meter, Yokogawa WT310. The maximum sampling speed of WattsUp Pro power meters is one sample every second. The accuracy specified in the data-sheets is $\pm{3\%}$. The minimum measurable power is 0.5 watts. The accuracy at 0.5 watts is $\pm{0.3}$ watts. The accuracy of Yokogawa WT310 is 0.1\% and the sampling rate is 100k samples per second.

HCLWattsUp API \cite{HCLWattsUpAPI} is used to gather the readings from the power meter to determine the dynamic energy consumption during the execution of PMMTG and PFFTTG applications. HCLWattsUp has no extra overhead and therefore does not influence the energy consumption of the application execution.

Fans are significant contributors to energy consumption. On our platform, fans are controlled in two zones: a) zone 0: CPU or System fans,  b) zone 1: Peripheral zone fans. There are 4 levels to control the speed of fans:
\begin{itemize}
	\item \emph{Standard}: BMC control of both fan zones, with CPU zone based on CPU temp (target speed 50\%) and Peripheral zone based on PCH temp (target speed 50\%)
	\item \emph{Optimal}: BMC control of the CPU zone (target speed 30\%), with Peripheral zone fixed at low speed (fixed ~30\%)
	\item \emph{Heavy IO}: BMC control of CPU zone (target speed 50\%), Peripheral zone fixed at 75\%
	\item \emph{Full}: all fans running at 100\%
\end{itemize}

To rule out the contribution of fans in dynamic energy consumption, we set the fans at full speed before executing the applications. When set at full speed, the fans run constantly at $ {\sim}13400 $ rpm until they are set to a different speed level. In this way, energy consumption due to fans is included only in the static power consumption of the platform. The temperature of our platform and speeds of the fans (with \emph{Full} setting) is monitored with the help of Intelligent Platform Management Interface (IPMI) sensors, both with and without the application run. An insignificant difference in the speeds of fans is found in both the scenarios.

\subsection{Parallel Matrix-Matrix Multiplication Using OpenBLAS DGEMM and Intel MKL DGEMM}

\subsubsection{Performance Optimization on a Single Socket Multicore CPU}

\begin{figure}
	\includegraphics[width=1\linewidth]{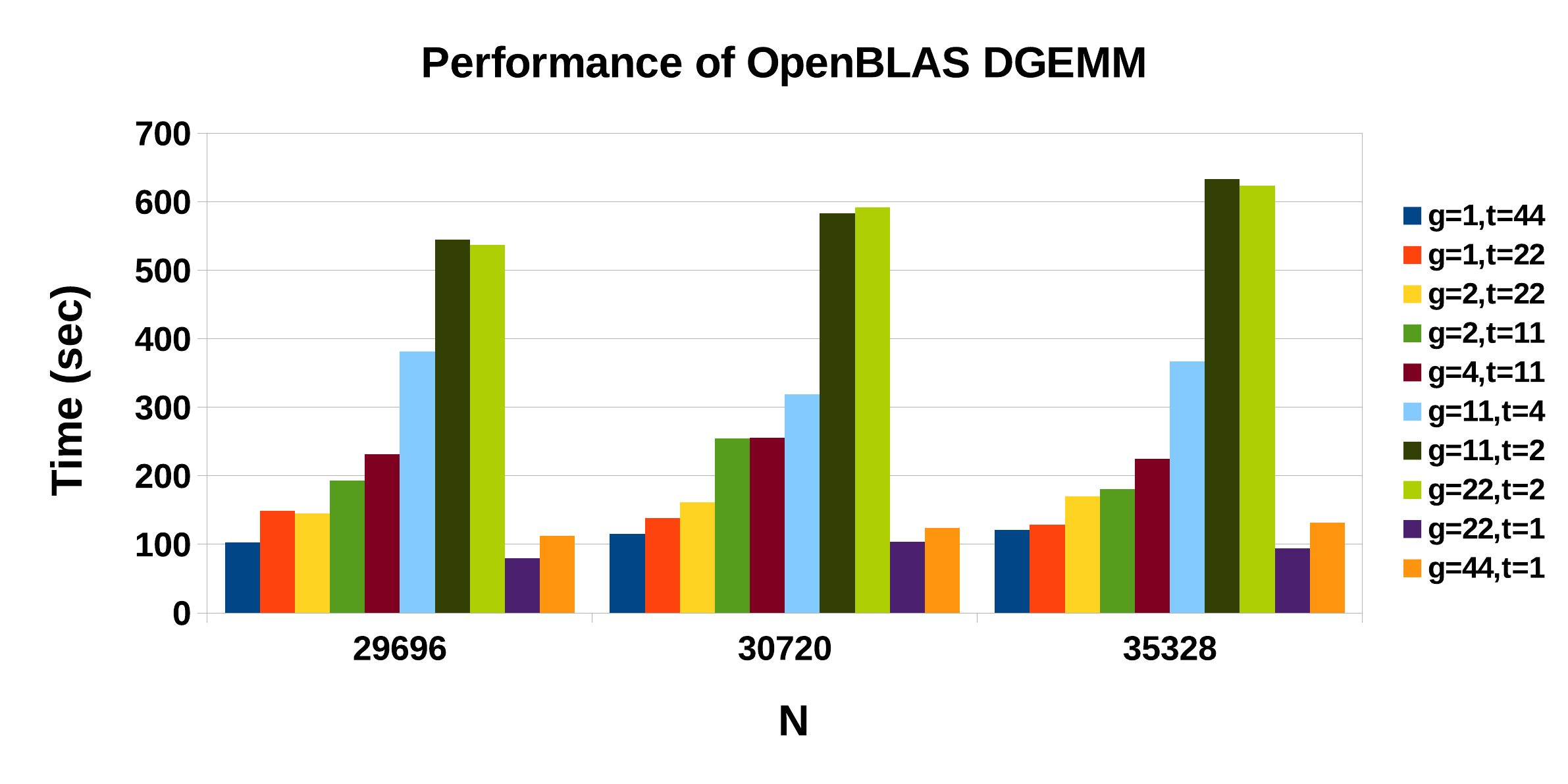}
	\caption{Performance of PMMTG application employing OpenBLAS DGEMM for different ($g$,$t$) configurations on S1.}
	\label{fig:perf_openblas_s1}
\end{figure}

Fiqure \ref{fig:perf_openblas_s1} shows the execution times of PMMTG using OpenBLAS DGEMM for different threadgroup combinations on a single-socket CPU (S1). The base version corresponds to the application configuration employing one threadgroup with optimal number of threads, which is 44 threads. The best combination is ($g$,$t$)=(22,1) for all the three workload sizes. It outperforms the base combination by 20\% for N=29696 and N=35328, and about 11\% for N=30720. Furthermore, the average performance improvement over the base combination for 41 tested workload sizes in the range, $5120 \leq N \leq 36000$, is 7\%. The starting problem size of 5120 is chosen to ensure that the workload size exceeds the last level cache.

\begin{figure}
	\includegraphics[width=1\linewidth]{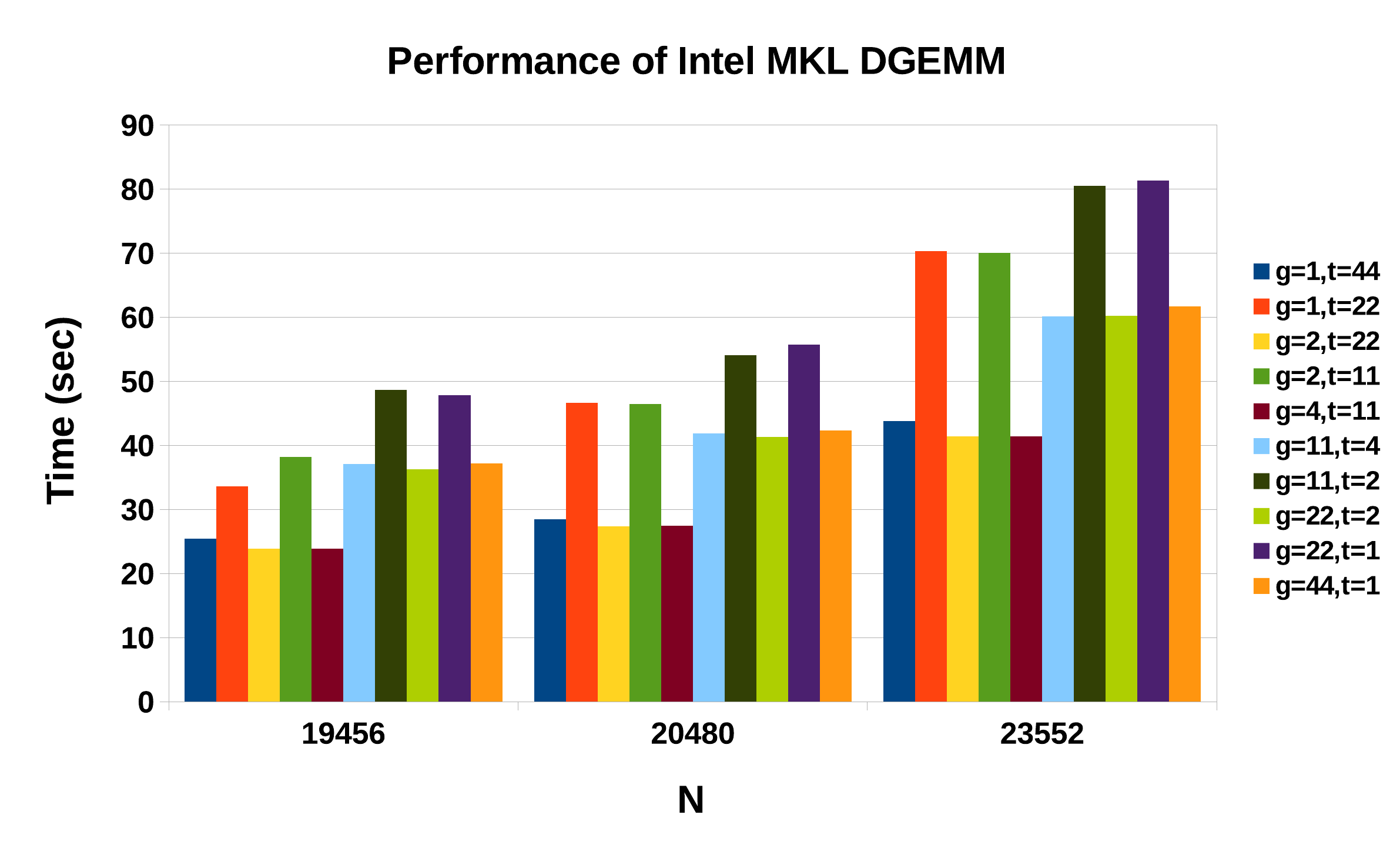}
	\caption{Performance of PMMTG application employing Intel MKL DGEMM for different ($g$,$t$) configurations on S1.}
	\label{fig:perf_inteldgemm_s1}
\end{figure}

Fiqure \ref{fig:perf_inteldgemm_s1} shows the execution times of PMMTG using Intel MKL DGEMM. The best combinations ($g$,$t$) are \{(4,11),(2,22)\} for all the three workload sizes. They outperform the base combination by 6\%. The average performance improvement over the base combination for 21 tested workload sizes in the range, $5120 \leq N \leq 36000$, is 5\%.

\subsubsection{Performance Optimization on a Dual-socket Multicore CPU}

\begin{figure}[!t]
\centering
\includegraphics[width=1\linewidth]{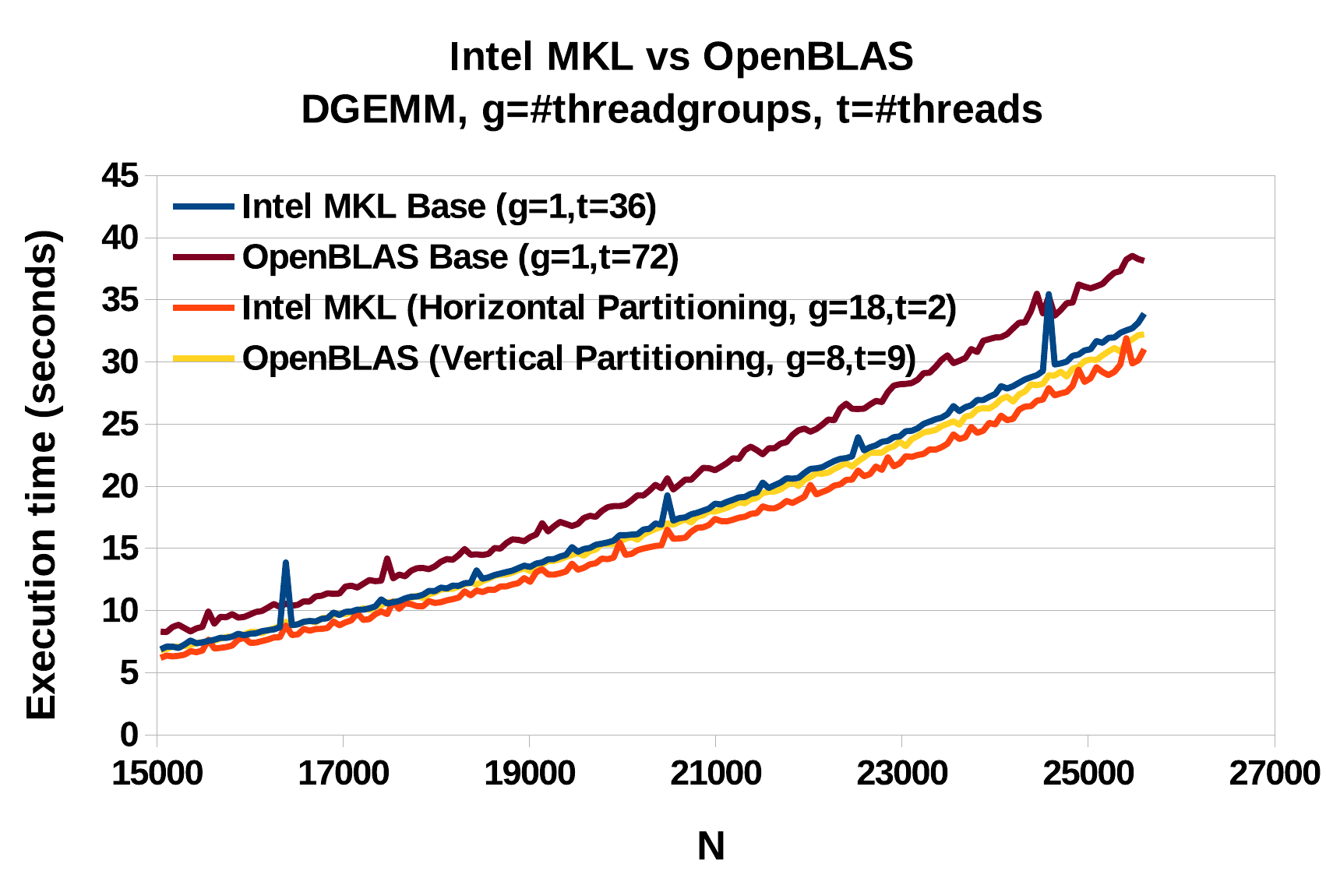}
\label{fig:intelmkl_openblas_dgemm_best}
\caption{Comparision between the base and best versions for Intel MKL DGEMM and OpenBLAS DGEMM on S3.}
\end{figure}

Figure \ref{fig:intelmkl_openblas_dgemm_best} shows the comparision between base and best combinations for OpenBLAS DGEMM and Intel MKL DGEMM on S3. The base version corresponds to application configuration employing one threadgroup with optimal number of threads.

Unlike the base version, the best combinations for OpenBLAS DGEMM and Intel MKL DGEMM do not have any performance variations (drops). The best combination for Intel MKL DGEMM is 18 threadgroups with 2 threads each. It outperforms the base version by 8\% on the average and the next best combination, 12 threadgroups with 2 threads each, by 2.5\%. Our solution method removed noticeable drops in performance for workload sizes 16384, 20480, and 24576, with performance improvements of 36.5\%, 14.5\% and 21.5\%.

\subsubsection{Energy Optimization on a Single Socket Multicore CPU}

\begin{figure}
	\includegraphics[width=1\linewidth]{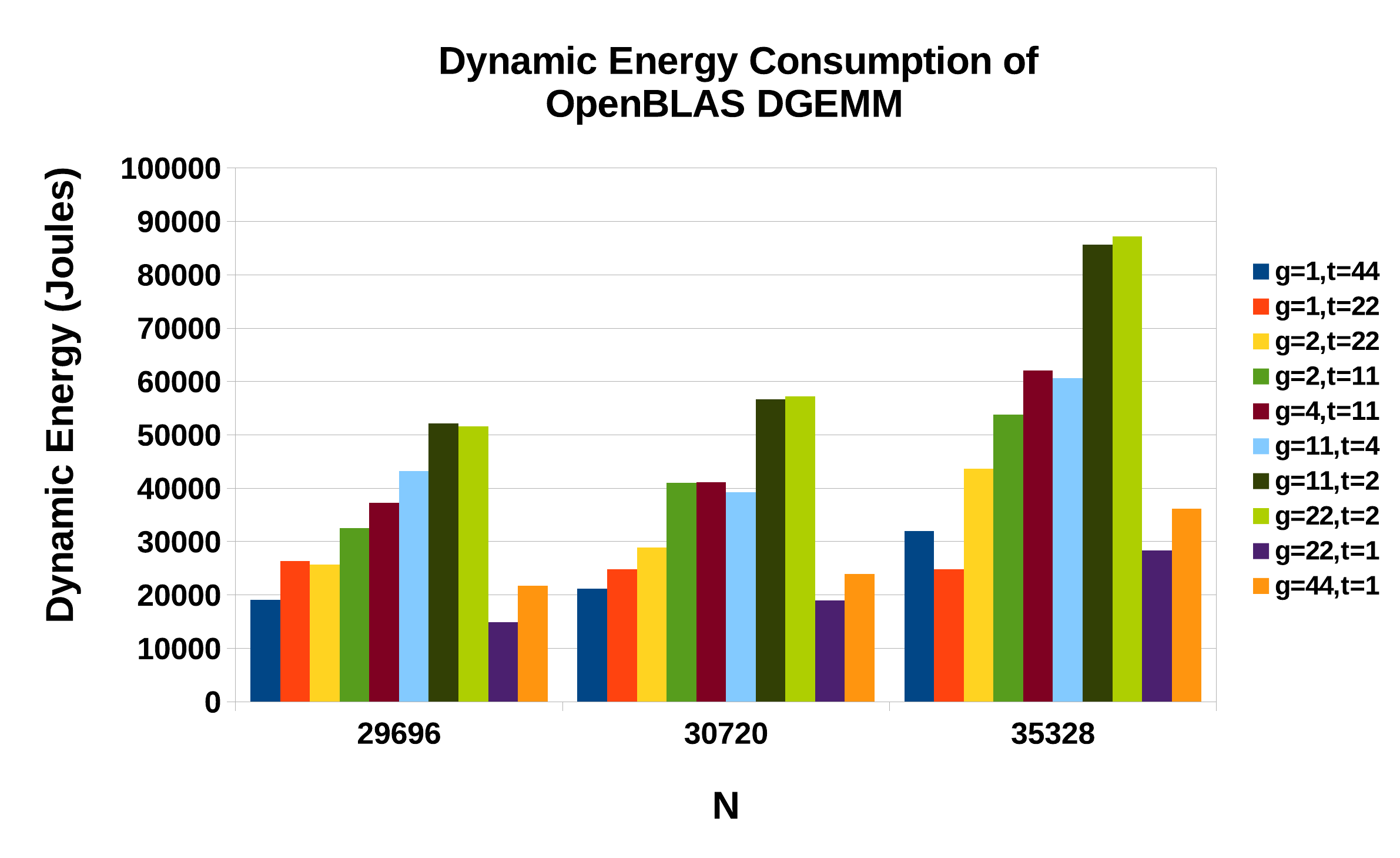}
	\caption{Dynamic energy consumption for PMMTG employing OpenBLAS DGEMM for different ($g$,$t$) configurations on S1.}
	\label{fig:dynamic_openblas_s1}
\end{figure}

Fiqure \ref{fig:dynamic_openblas_s1} shows the dynamic energy consumptions for PMMTG using OpenBLAS DGEMM of different threadgroup combinations on a single-socket CPU (S1). The base version corresponds to application configuration employing one threadgroup with optimal number of threads, which is 44 threads. The best combination for sizes N=29696 and N=30720 is ($g$,$t$)=(22,1). It outperforms the base combination by 20\%. The best combination for $N=35328$ is ($g$,$t$)=(1,22), which outperforms the base combination by 23\%. Furthermore, the average improvement (or energy savings) over the base combination for 41 tested workload sizes in the range, $5120 \leq N \leq 35000$, is 8\%. 

\begin{figure}
	\includegraphics[width=1\linewidth]{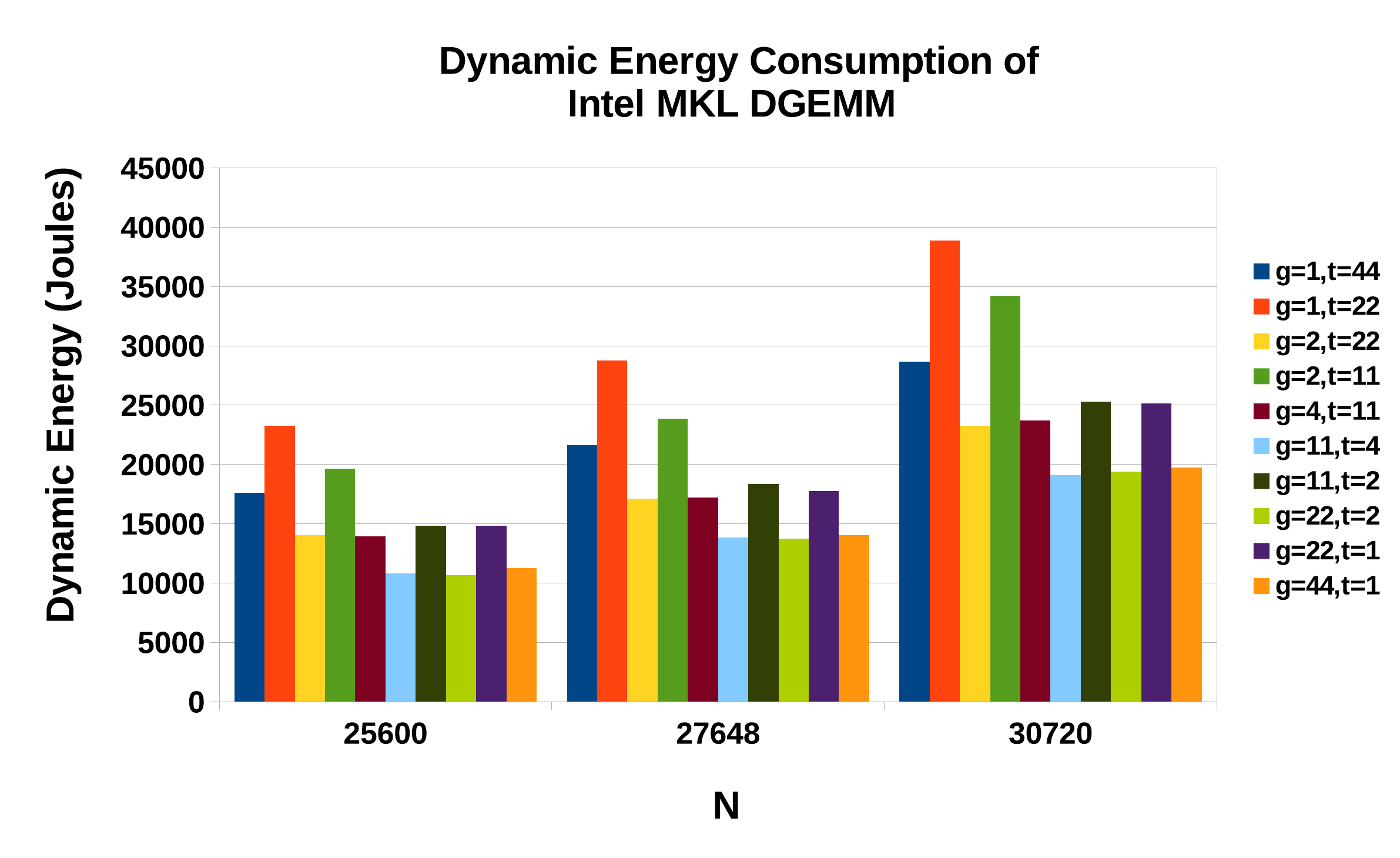}
	\caption{Dynamic energy consumption for PMMTG employing Intel MKL DGEMM for different ($g$,$t$) configurations on S1.}
	\label{fig:dynamic_inteldgemm_s1}
\end{figure}

Fiqure \ref{fig:dynamic_inteldgemm_s1} shows the dynamic energy consumptions for PMMTG using Intel MKL DGEMM. There are three best combinations for each problem size, ($g$,$t$)=\{(11,4),(22,2),(44,1)\}. They outperform the base combination by 35\%. Furthermore, the average improvement over the base combination for 21 tested workload sizes in the range, $5120 \leq N \leq 35000$, is 35.7\%.

\subsubsection{Energy Optimization on a Dual-socket Multicore CPU}

\begin{figure}
	\includegraphics[width=1\linewidth]{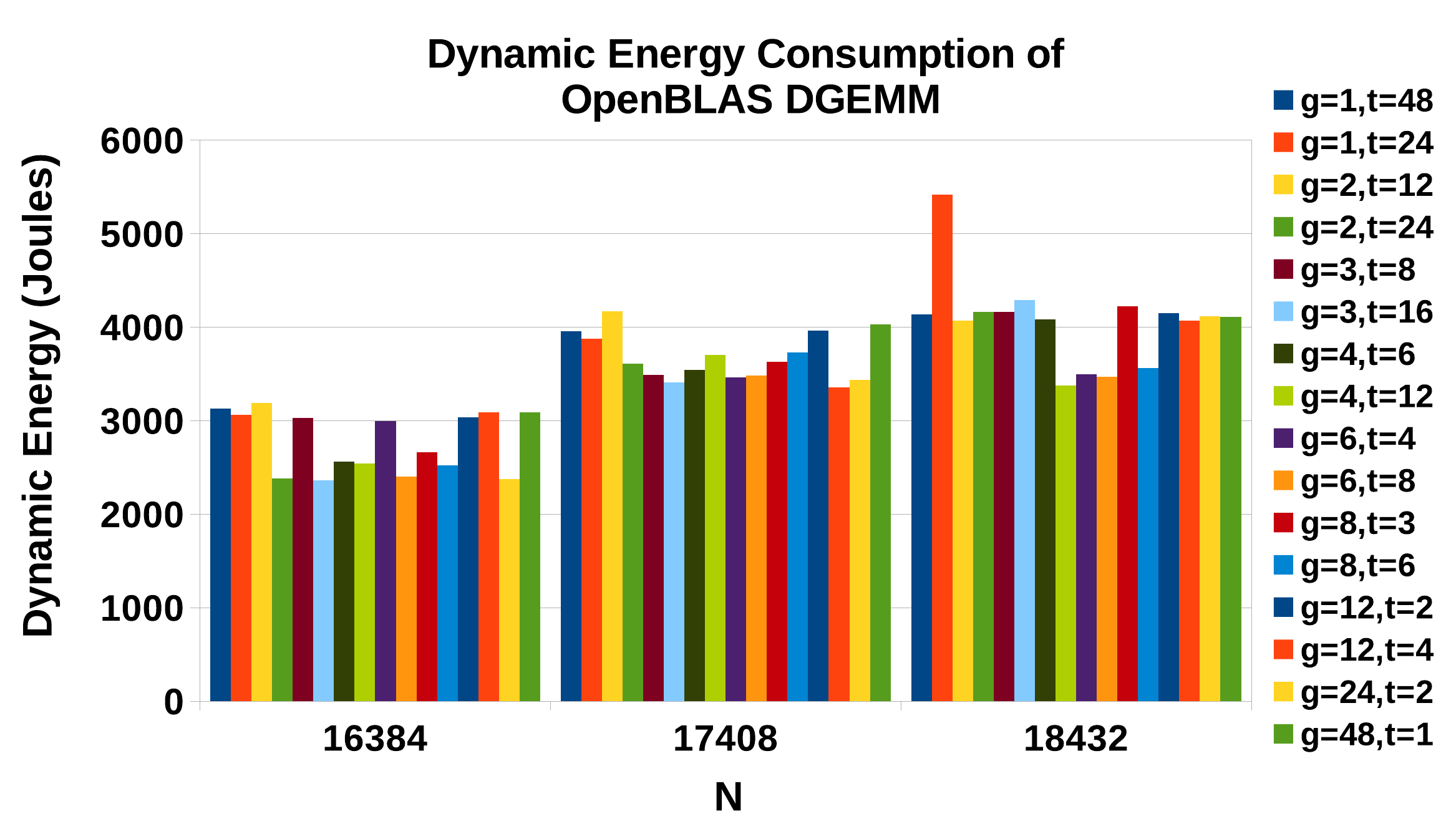}
	\caption{Dynamic energy consumption of PMMTG employing OpenBLAS DGEMM for different ($g$,$t$) configurations on S2.}
	\label{fig:dynamic_openblas}
\end{figure}

Figure \ref{fig:dynamic_openblas} show the results for PMMTG based on OpenBLAS DGEMM on S2 with three different workload sizes. There are four best combinations minimizing the dynamic energy consumption for workload size 16384, ($g$,$t$)=\{(2,24),(3,16),(6,8),(24,2)\}. The energy savings for these combinations compared with the best base combination, ($g$,$t$)=(1,24), is around 21\%. For the workload sizes 17408 and 18432, the best combinations are (12,4) and (4,12). The energy savings in comparison with the best base combination, ($g$,$t$)=(1,24), for 17408 and ($g$,$t$)=(1,44) for 18432, are 15\% and 18\%. Furthermore, the average improvement over the best base combination for 19 tested workload sizes in the range, $5120 \leq N \leq 35000$, is 10\%.

\begin{figure}
	\includegraphics[width=1\linewidth]{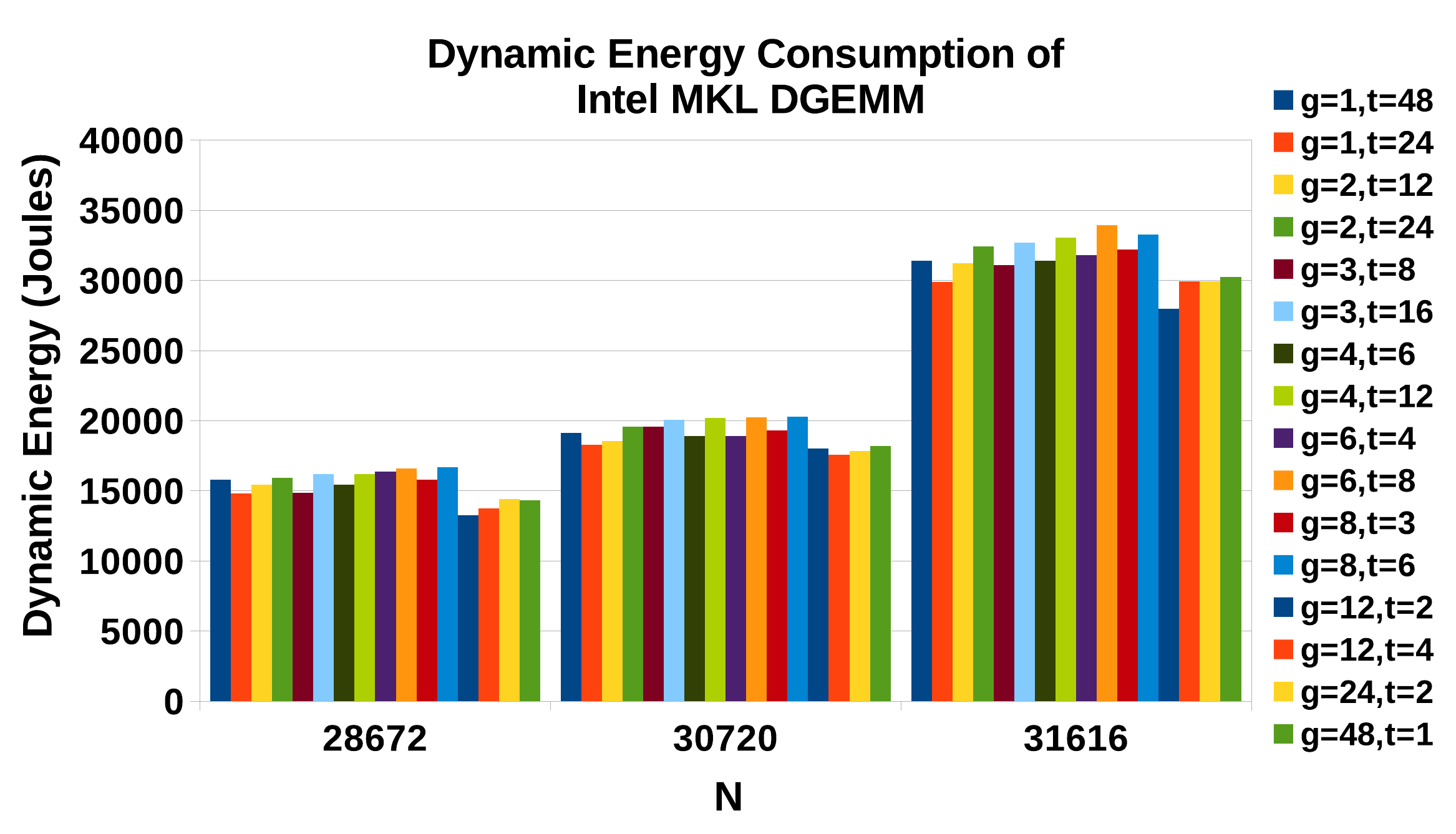}
	\caption{Dynamic energy consumption of PMMTG employing Intel MKL DGEMM for different ($g$,$t$) configurations on S2.}
	\label{fig:dynamic_inteldgemm}
\end{figure}

Figure \ref{fig:dynamic_inteldgemm} show the results for PMMTG based on Intel MKL DGEMM on S2. The best combination minimizing the dynamic energy consumption for workload size 28672 involves 12 threadgroups with 2 threads each. The energy savings for this combination compared with the best base combination, (1,24), is 10.5\%. For the workload sizes 30720 and 31616, the best combinations are (12,4) and (12,2). The energy savings in comparison with the best base combination are 4\% and 7\%. Furthermore, the average improvement over the best base combination for 19 tested workload sizes in the range, $5120 \leq N \leq 35000$, is 13\%.

\subsection{Parallel 2D Fast Fourier Transform Using FFTW and Intel MKL FFT}

In this section, we use 2D fast Fourier transform routines from two packages, FFTW-3.3.7 and Intel MKL. The packages are installed with multithreading, SSE/SSE2, AVX2, and FMA (fused multiply-add) optimizations enabled. For Intel MKL FFT, no special environment variables are used.  Three planner flags, \{FFTW\_ESTIMATE, FFTW\_MEASURE, FFTW\_PATIENT\} were tested. The execution times for the flags \{FFTW\_MEASURE, FFTW\_PATIENT\} are high compared to those for FFTW\_ESTIMATE. The long execution times are due to the lengthy times to create the plans because FFTW\_MEASURE tries to find an optimized plan by computing many FFTs whereas FFTW\_PATIENT considers a wider range of algorithms to find a more optimal plan.

\subsubsection{Performance Optimization on a Single Socket Multicore CPU}

\begin{figure}
	\includegraphics[width=1\linewidth]{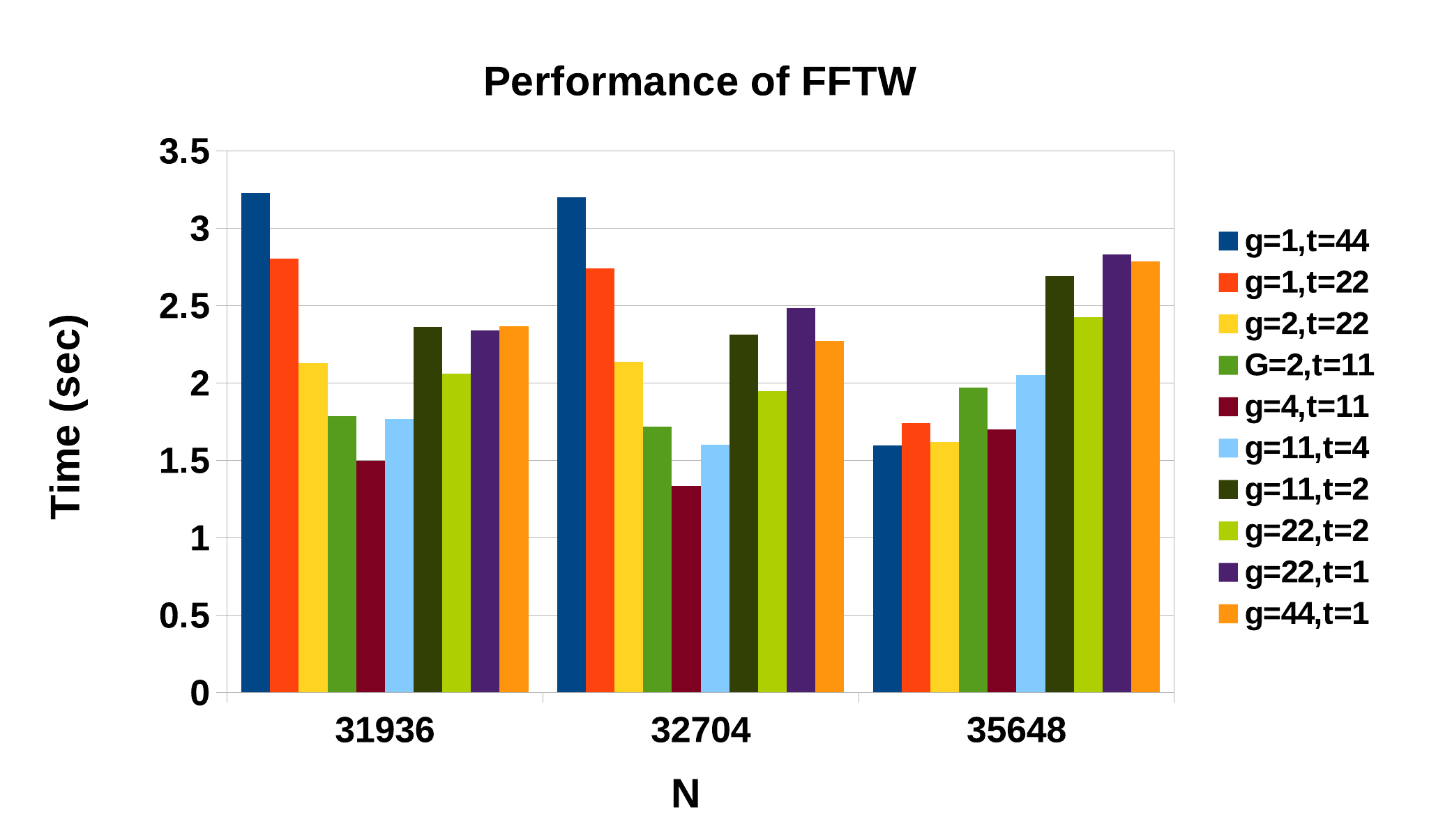}
	\caption{Performance of PFFTTG employing FFTW for different ($g$,$t$) configurations on S1.}
	\label{fig:fftw_s1}
\end{figure}

Figure \ref{fig:fftw_s1} shows the results for PFFTTG employing FFTW on a single-socket CPU (S1). The best combination, ($g,t$)=(4,11), is the same for workload sizes, N=31936 and N=32704. The improvements over the base combination, ($g,t$)=(1,44), are 55\% and 57\%. For matrix dimension, N=35648, the base combination is the best and outperforms the next best combination, ($g,t$)=(2,22), by 5\%.

\begin{figure}
	\includegraphics[width=1\linewidth]{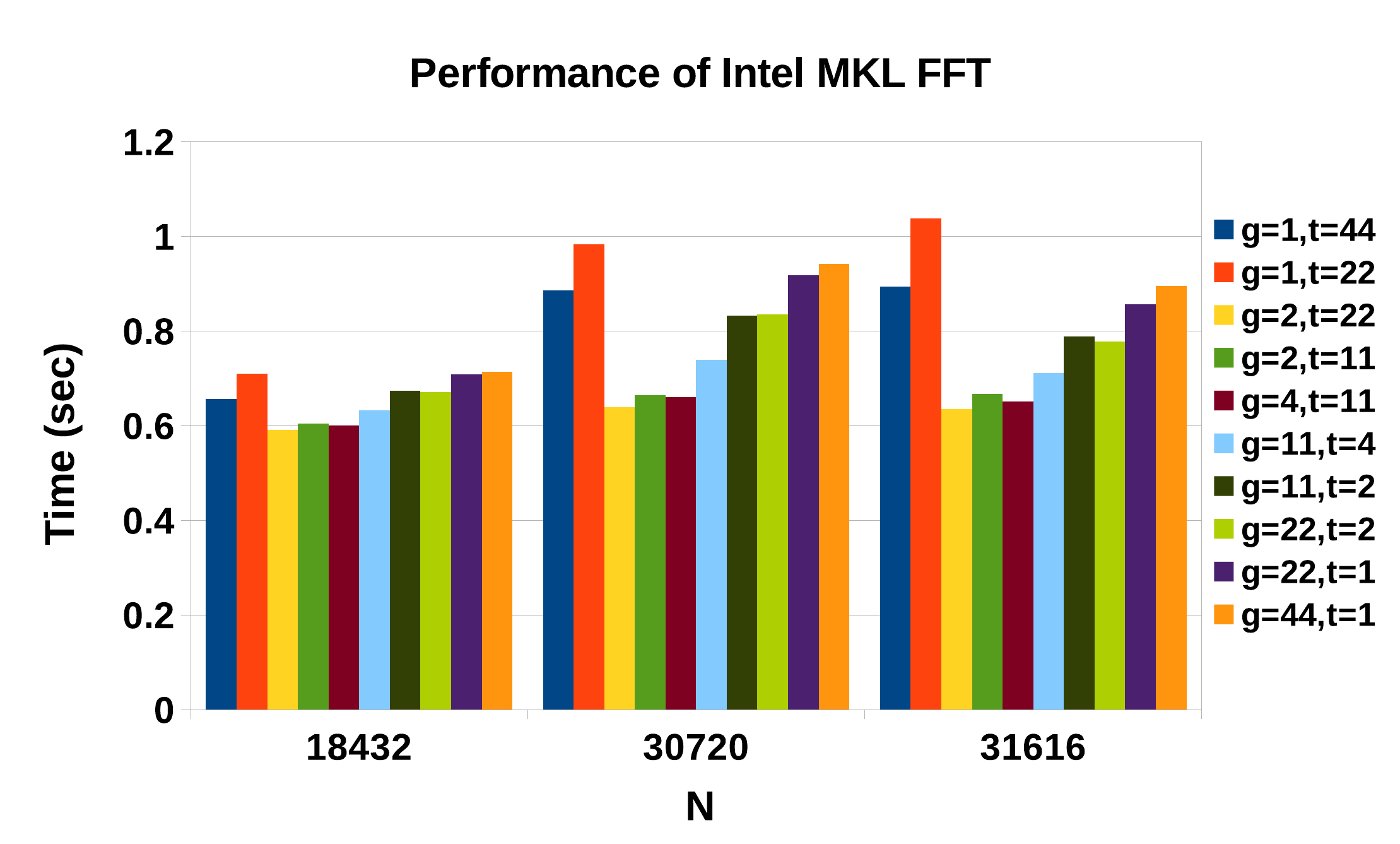}
	\caption{Performance of PFFTTG employing Intel MKL FFT for different ($g$,$t$) configurations on S1.}
	\label{fig:perf_intelfft_s1}
\end{figure}

Figure \ref{fig:perf_intelfft_s1} shows the results for PFFTTG employing Intel MKL FFT. There are three best combinations, ($g,t$)=(2,22),(2,11),(4,11), for all the three workload sizes, where performances differ from each other by less than 5\%. Their improvement over the base combination, ($g,t$)=(1,44), for N=18432 is 8\%. For workload sizes, N=30720 and N=31616, the performance improvements are 25\% and 26\%. Furthermore, the average performance improvement over the best base combination for 23 tested workload sizes in the range, $5120 \leq N \leq 37000$, is 27\%. 

\subsubsection{Performance Optimization on Dual-socket Multicore CPUs}

All results in this section are represented by a 3D surface represented by axes for performance or energy, number of threadgroups ($g$) and the number of threads in each threadgroup, $t$. The location of the minimum in the surface is shown by the red dot.

\begin{figure}[!t]
\centering
\subfloat[][]{
\includegraphics[width=1\linewidth]{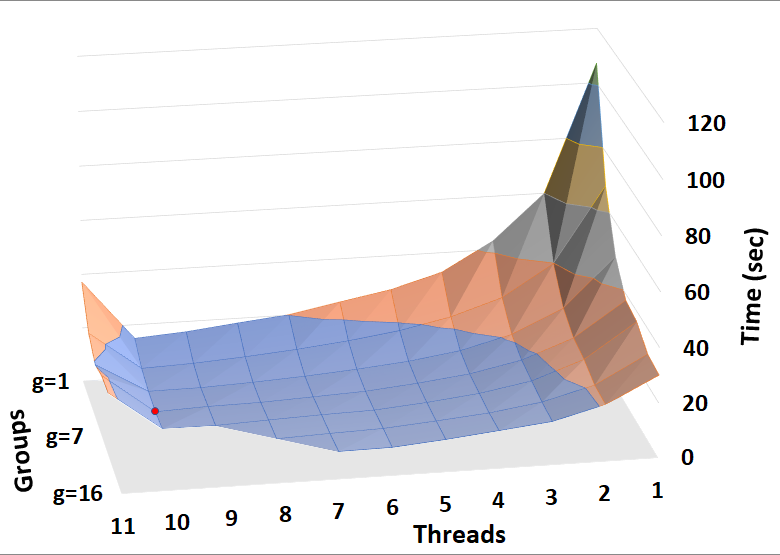}
\label{fig:30976_56}}
\hfill
\subfloat[][]{
\includegraphics[width=1\linewidth]{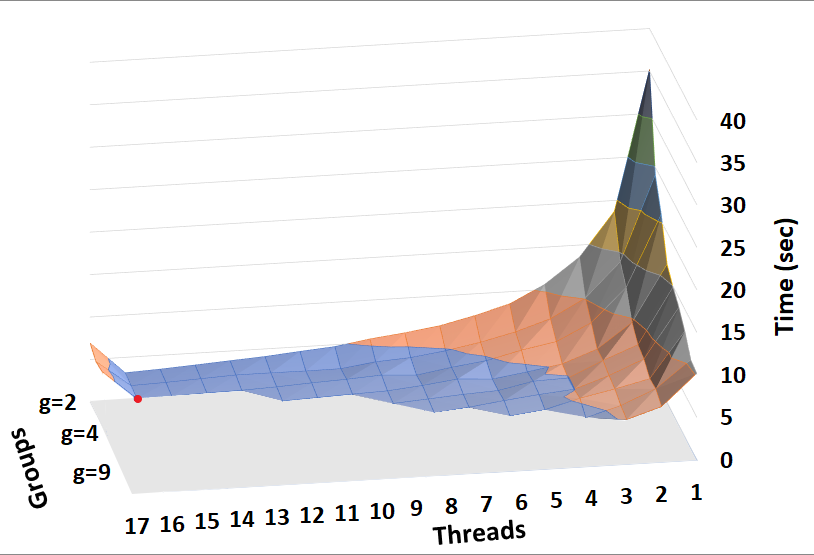}
\label{fig:17728_36}}
\hfill
\caption{(a). Performance profile of FFTW PFFTTG for different ($g$,$t$) configurations on S4 for workload size, N=30976. (b). Performance profile of FFTW PFFTTG for different ($g$,$t$) configurations on S3 for workload size, N=17728. Red dot represents the minimum.}
\end{figure}

Figure \ref{fig:30976_56} shows the results of PFFTTG using FFTW3.3.7 on S4 for matrix dimension N=30976. The area with minimum execution time is located in the figure in the region containing \{4,7,8\} threadgroups with 10 threads in each group. The minimum is achieved for the combination ($g,t$)=(7,10) with the execution time of 8 seconds. The speedup is around 100\% in comparison with the best combination of threads for one group ($g,t$)=(1,10) where the execution time is 16 seconds. 

Figure \ref{fig:17728_36} presents the results of PFFTTG using FFTW3.3.7 on S3 for the matrix dimension N=17728. The minimum is centred around number of threadsgroups equal to \{4,7,8\}. The minimum is achieved for the combination, ($g,t$)=(4,16). The performance improvement is 80\% in comparison with ($g,t$)=(1,72), which is the best combination for one group.

\subsubsection{Energy Optimization on a Single Socket Multicore CPU}

\begin{figure}
	\includegraphics[width=1\linewidth]{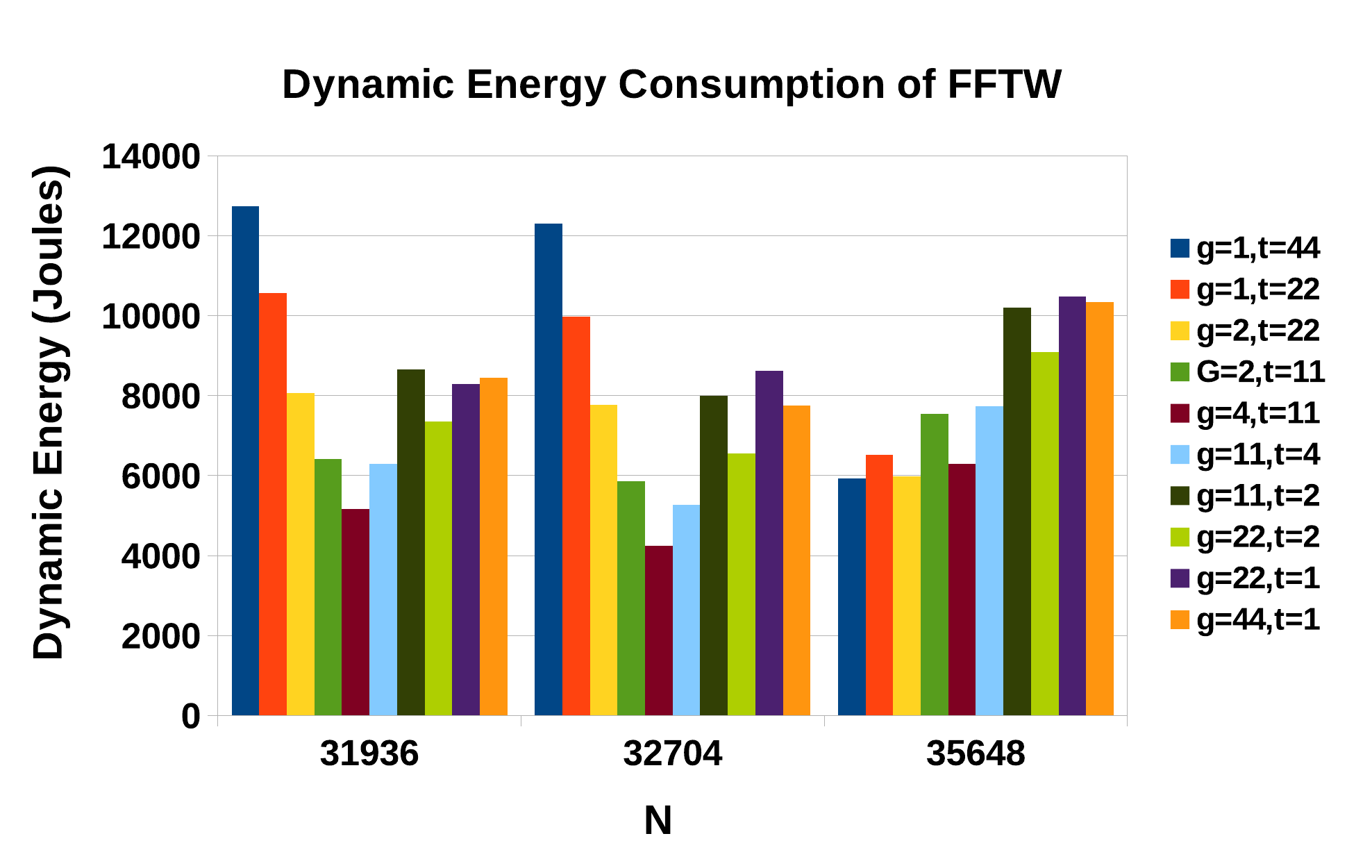}
	\caption{Dynamic energy consumption of PFFTTG employing FFTW for different ($g$,$t$) configurations on S1.}
	\label{fig:fftw_dynamic_s1}
\end{figure}

Figure \ref{fig:fftw_dynamic_s1} shows the dynamic energy comparision for PFFTTG employing FFTW between base and best combinations for workload sizes, 31936, 32704, and 35648 on a single-socket CPU (S1). The best combination ($g,t$)=(4,11) is the same for workload sizes, 31936 and 32704. The reductions in dynamic energy consumption in comparison with the base combination, ($g,t$)=(1,44), are 41\% and 65\%. For workload size 35648, the base combination is the best and outperforms the next best combination ($g,t$)=(2,22) by 5\%. For Intel MKL FFT, the base combination, ($g$,$t$)=(1,44), is the best.

\subsubsection{Energy Optimization on a Dual-socket Multicore CPU}

\begin{figure}[!t]
\centering
\subfloat[][]{
\includegraphics[width=1\linewidth]{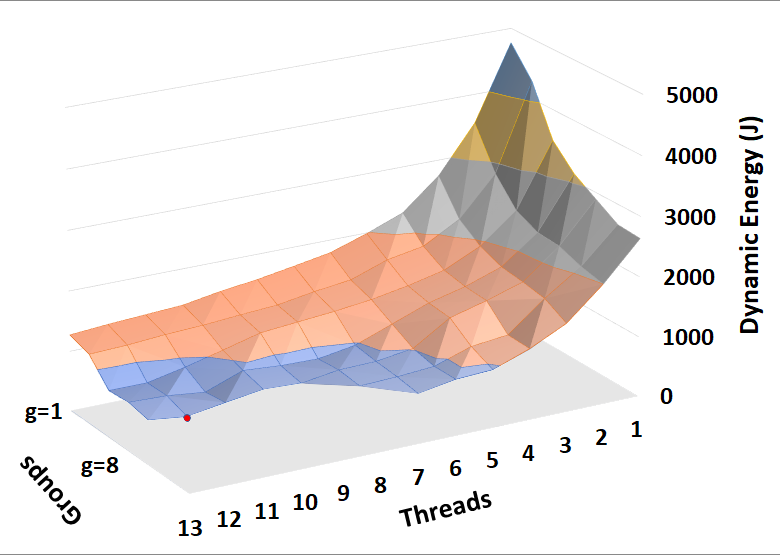}
\label{fig:30464_56_e}}
\hfill
\subfloat[][]{
\includegraphics[width=1\linewidth]{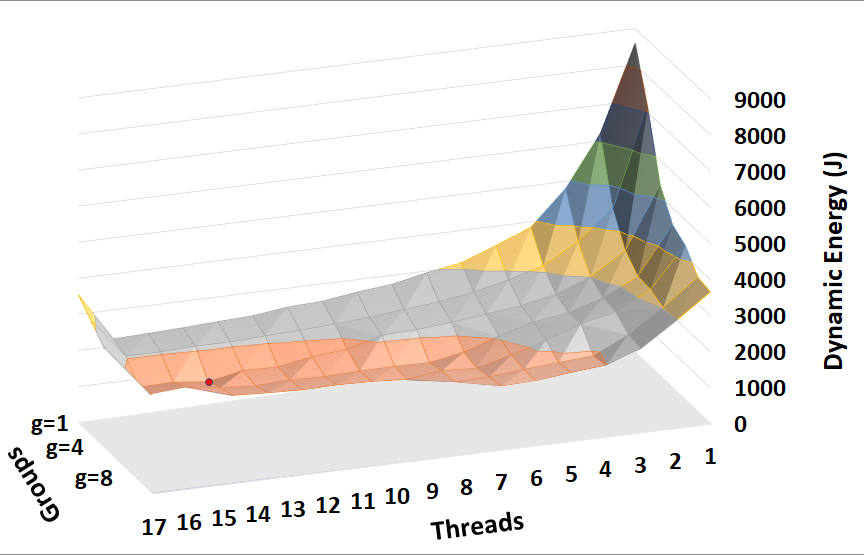}
\label{fig:32192_56_e}}
\hfill
\caption{(a). Energy profile of FFTW PFFTTG for different ($g$,$t$) configurations on S4 for workload size N=30464. (b). Energy profile of FFTW PFFTTG for different ($g$,$t$) configurations on S4 for workload size N=32192. Red dot represents the minimum.}
\end{figure}

Figures \ref{fig:30464_56_e}, \ref{fig:32192_56_e} show the results for PFFTTG employing FFTW on S4 for matrix sizes equal to N=30464 and N=32192. The minimum for dynamic energy is located in \{4,7,8\} threadgroups with 14 threads in each threadgroup for workload size (N=32192) and 12 threads in each threadgroup for workload size 30464. The minimum for the workload size 30464 is achieved for the combination, ($g,t$)=(8,12). The dynamic energy consumption for this combination is 661 Joules. The energy saving is around 30\% in comparison with the best combination of threads for one group ($g,t$)=(1,45) whose dynamic energy consumption is 918 Joules. The minimum for the workload size (N=32192) is achieved for the combination, ($g,t$)=(4,14). The saving is around 35\% in comparison with ($g,t$)=(1,16) where dynamic energy is 2197 Joules.

\subsection{Bi-Objective Optimization for Performance and Dynamic Energy}

\subsubsection{Single Socket Multicore CPU}

\begin{figure}[!t]
\centering
\subfloat[][]{
\includegraphics[width=1\linewidth]{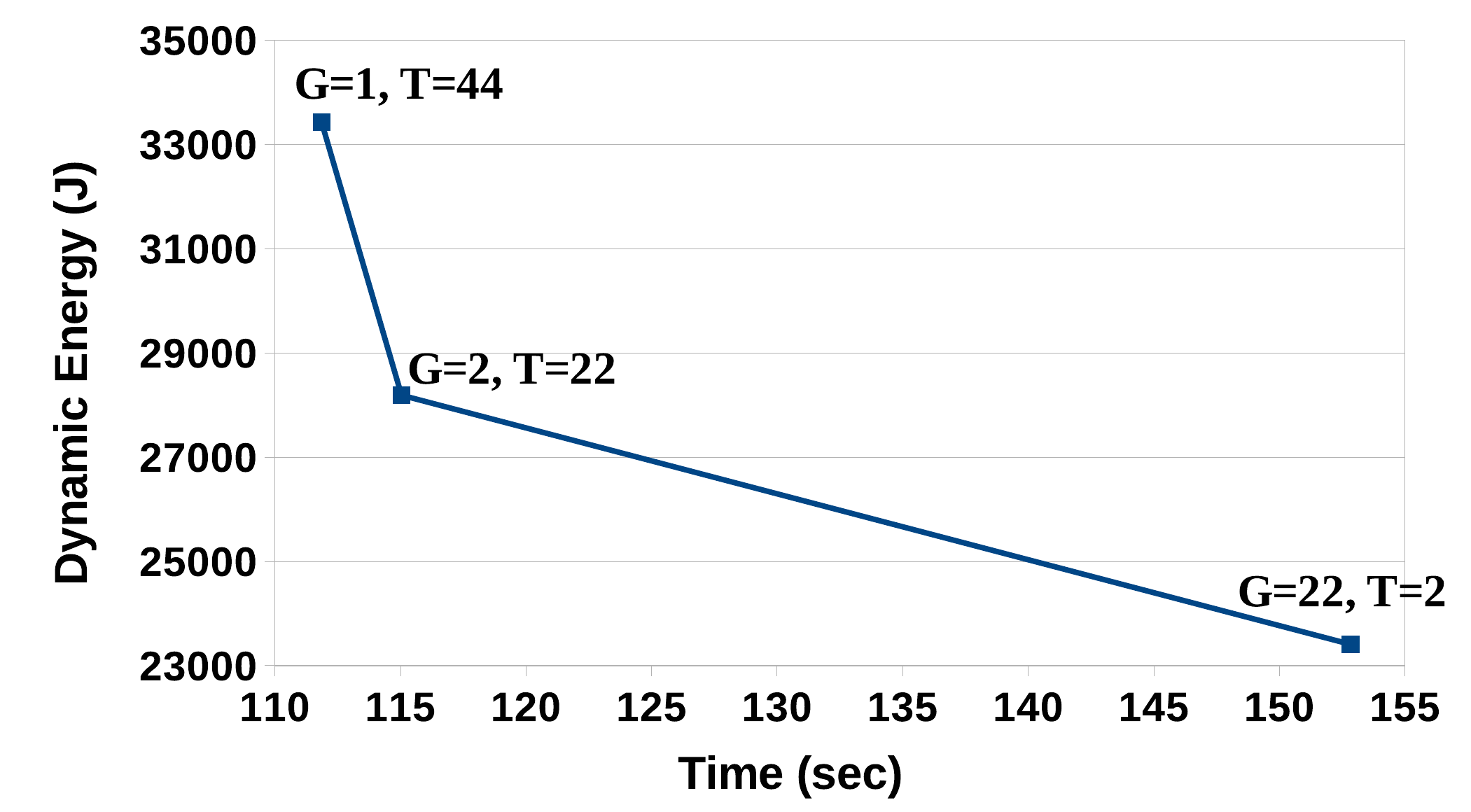}
\label{fig:energy_32768_s1}}
\hfill
\subfloat[][]{
\includegraphics[width=1\linewidth]{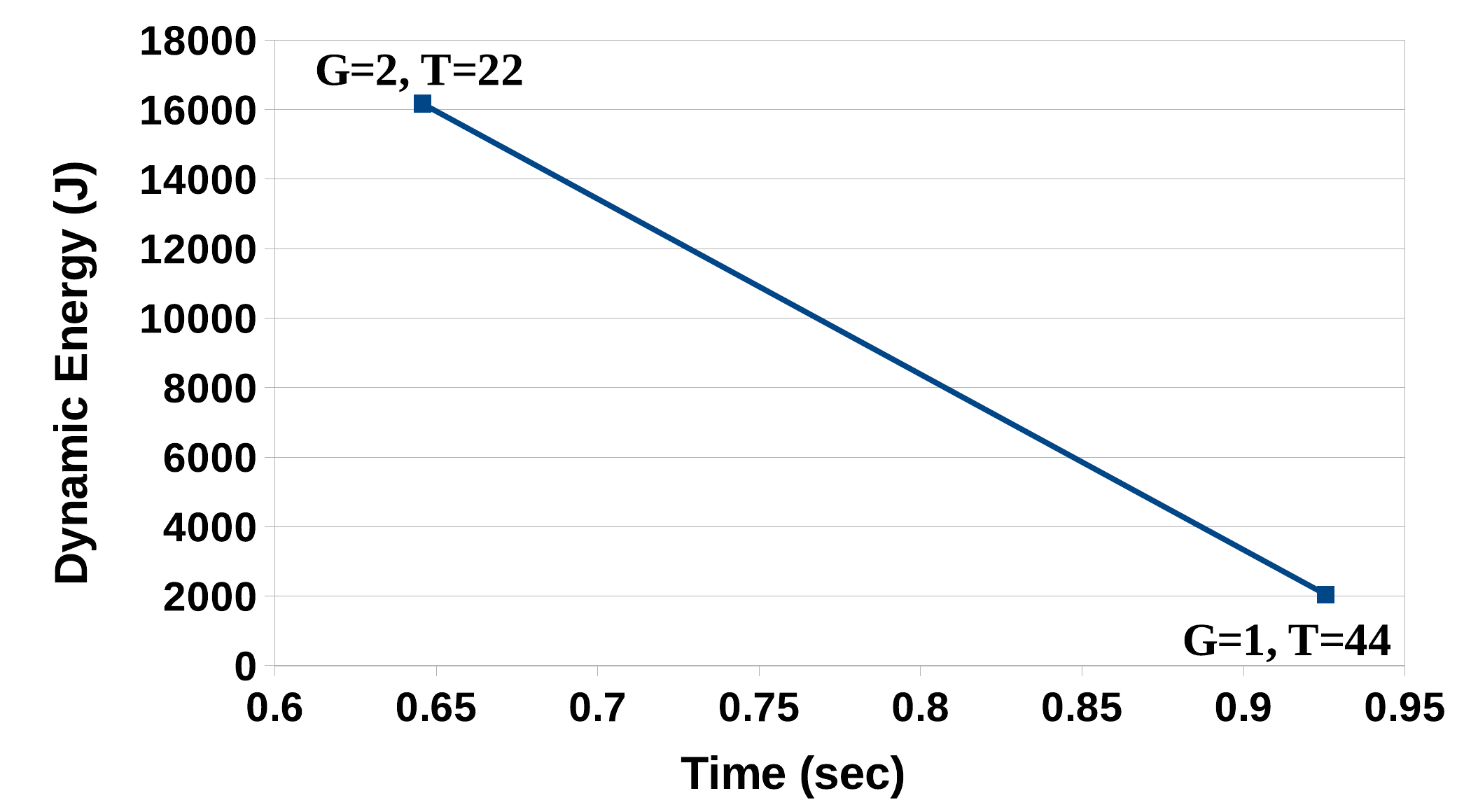}
\label{fig:energy_31744_s1}}
\hfill
\caption{(a). Pareto-optimal front of Intel MKL DGEMM PMMTG application on S1 for workload size N=32768. (b). Pareto-optimal front of Intel MKL FFT PFFTTG on S1 for workload size N=31744.}
\end{figure}

Figure \ref{fig:energy_32768_s1} shows the globally Pareto-optimal front for PMMTG employing Intel MKL DGEMM on S1 for workload size 32768. Optimizing for dynamic energy consumption alone degrades performance by 27\%, and optimizing for performance alone increases dynamic energy consumption by 30\%. The average and maximum sizes of the Pareto-optimal fronts for Intel MKL DGEMM are (2.3,3).

Figure \ref{fig:energy_31744_s1} shows the globally Pareto-optimal front for PFFTTG based on Intel MKL FFT on S1 for workload size 31744. There are two globally Pareto-optimal solutions. Optimizing for dynamic energy consumption alone degrades performance by around 31\%, and optimizing for performance alone increases dynamic energy consumption by 87\%. The average and maximum sizes of the Pareto-optimal fronts for Intel MKL FFT are (2.6,3).

No bi-objective trade-offs were observed for FFTW and OpenBLAS applications. We will investigate two lines of research in our future work. One is the influence of workload distribution; The other is the absence of bi-objective trade-offs for open-source packages such as FFTW and OpenBLAS using a dynamic energy predictive model.

\subsubsection{Dual-socket Multicore CPUs}

In this section, we will focus on bi-objective optimization on dual-socket CPUs, S2 and S4.

\begin{figure}[!t]
\centering
\subfloat[][]{
\includegraphics[width=1\linewidth]{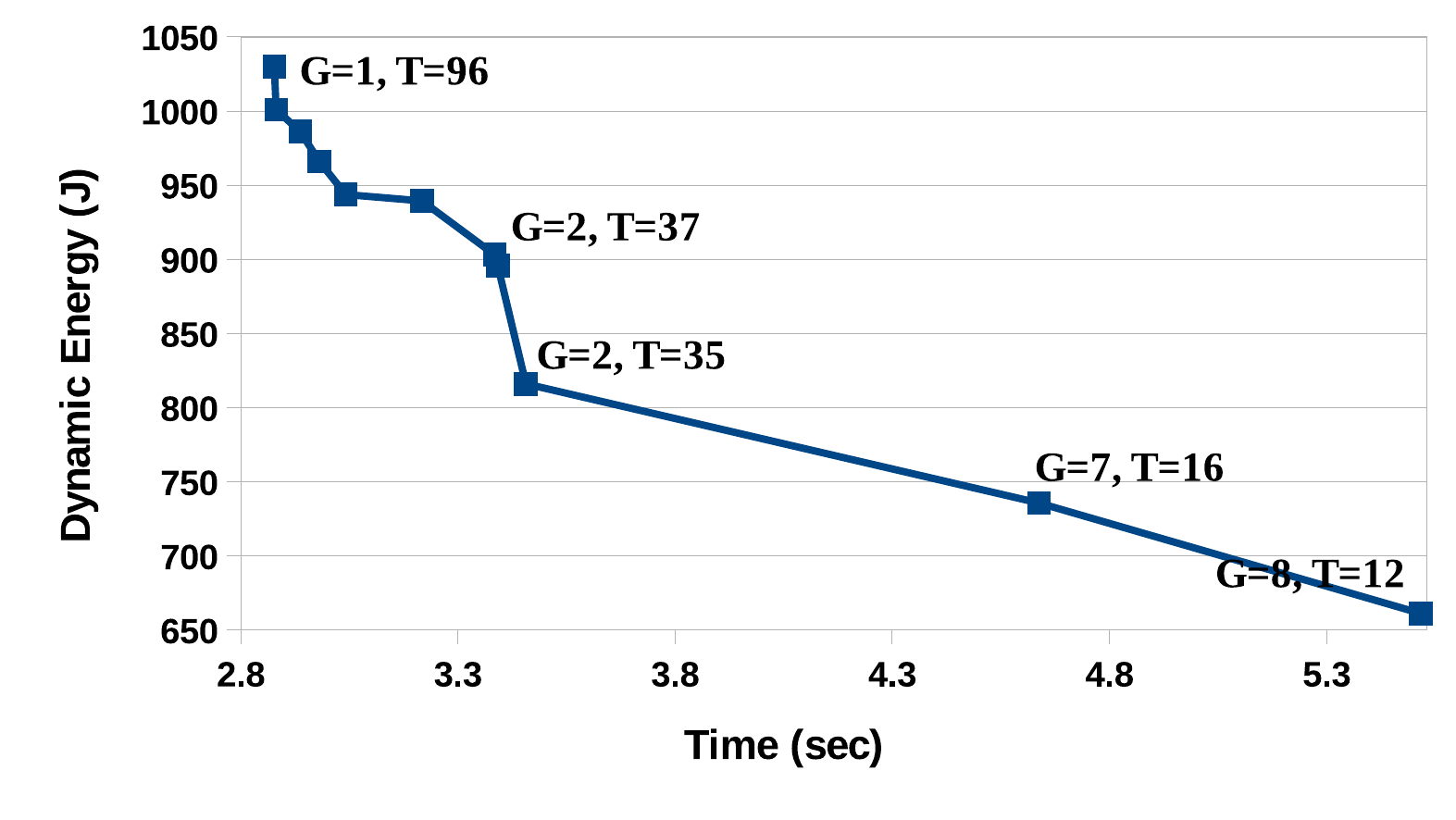}
\label{fig:energy_30464}}
\hfill
\subfloat[][]{
\includegraphics[width=1\linewidth]{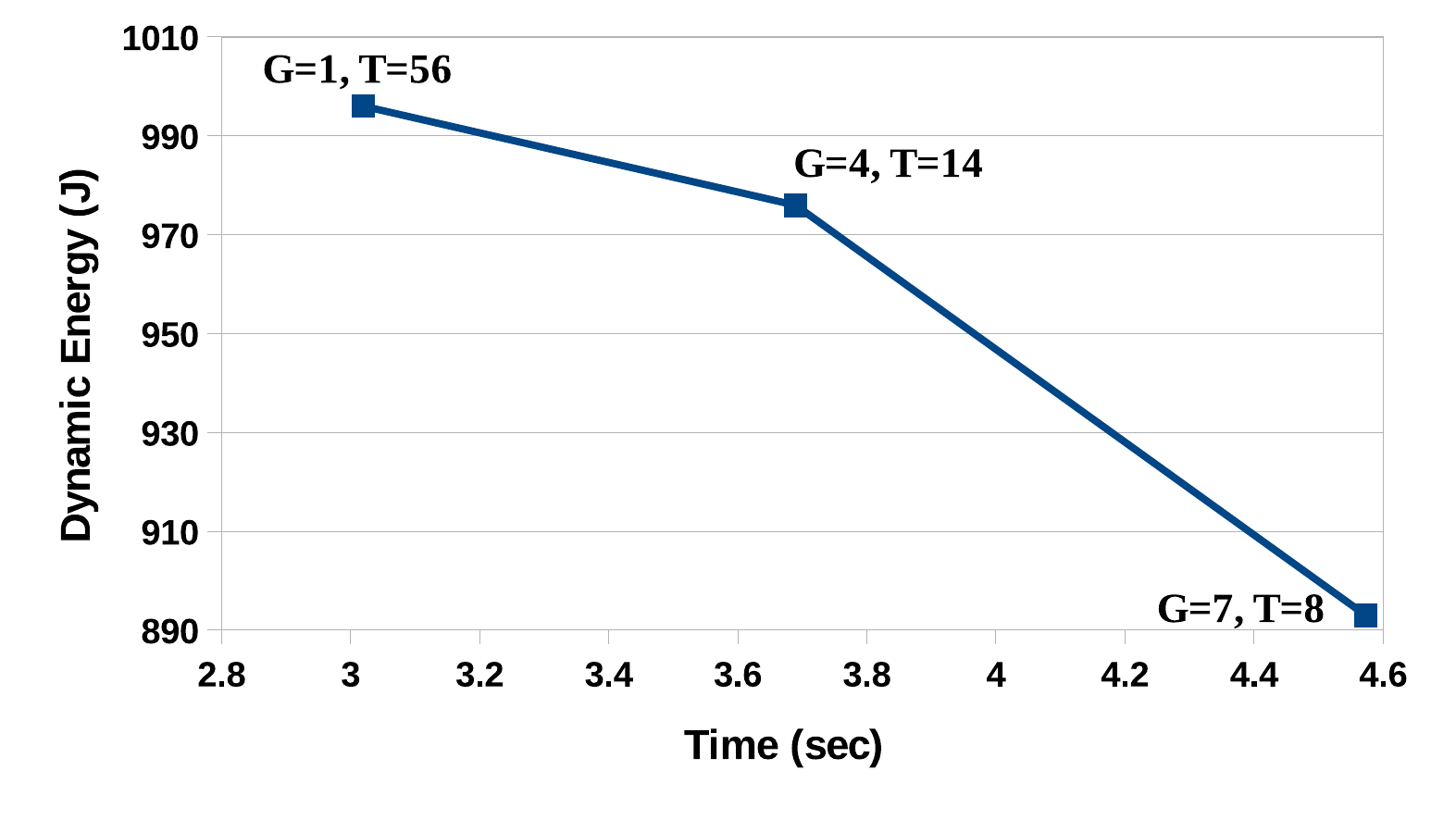}
\label{fig:energy_22208_intelfft}}
\hfill
\caption{(a). Pareto-optimal front of FFTW PFFTTG on S4 for workload size, N=30464. (b). Pareto-optimal front of Intel MKL FFT PFFTTG on S4 for workload size, N=22208.}
\end{figure}

Figures \ref{fig:energy_30464} shows the globally Pareto-optimal fronts for PFFTTG FFTW on S4 for workload size, N=30464. The maximum number of globally Pareto-optimal solutions is 11. The optimization for dynamic energy consumption alone degrades performance by 49\%, and optimizing for performance alone increases dynamic energy consumption by 35\%.

Figure \ref{fig:energy_22208_intelfft} shows the globally Pareto-optimal front for PFFTTG employing Intel MKL FFT on S2 for workload size, N=22208. Optimizing for dynamic energy consumption alone degrades performance by 33\%, and optimizing for performance alone increases dynamic energy consumption by 10\%. The average and maximum sizes of the Pareto-optimal fronts for FFTW and Intel MKL FFT are (3,11) and (2.7, 3).

\begin{figure}[!t]
\centering
\subfloat[][]{
\includegraphics[width=1\linewidth]{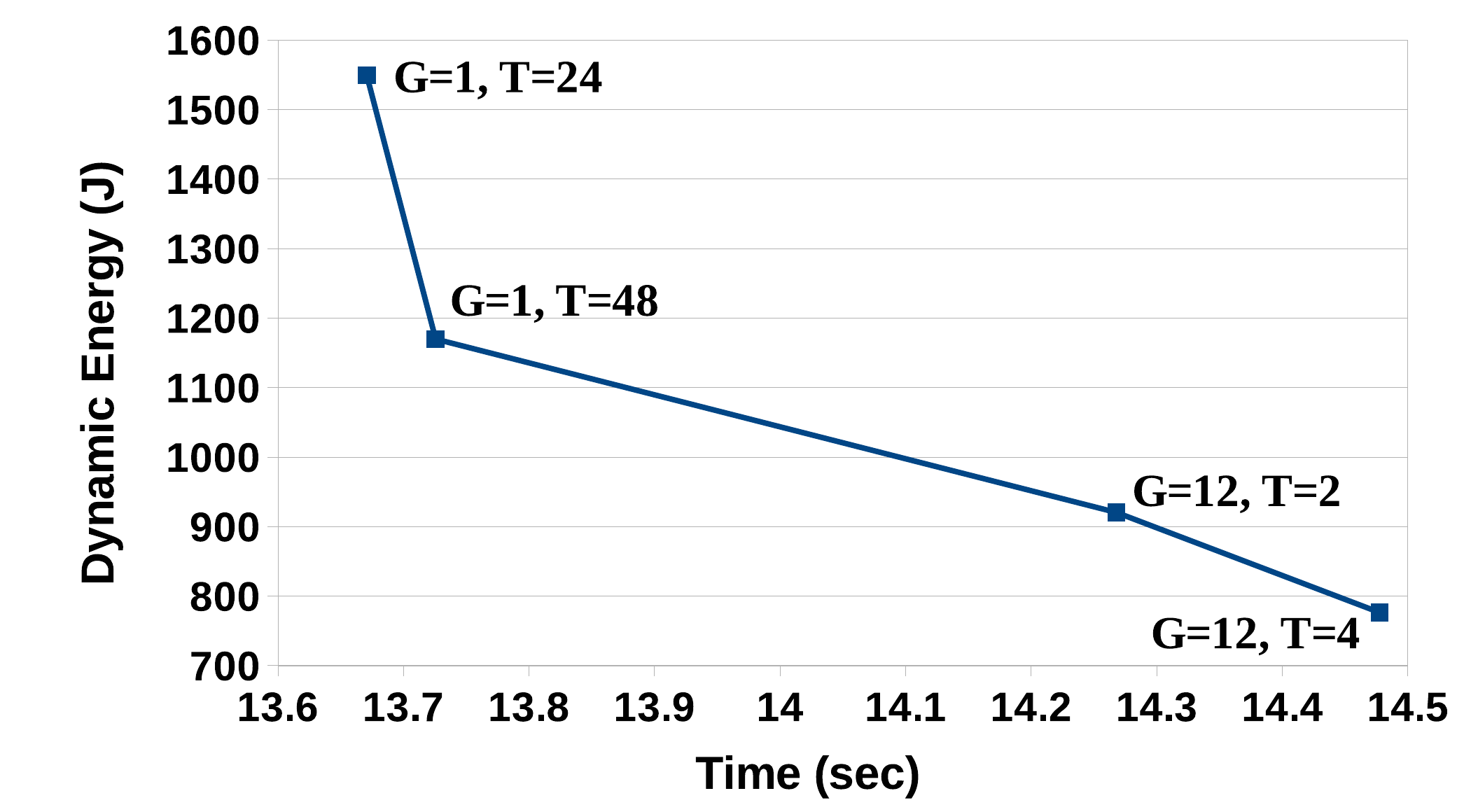}
\label{fig:energy_17408_intel}}
\hfill
\subfloat[][]{
\includegraphics[width=1\linewidth]{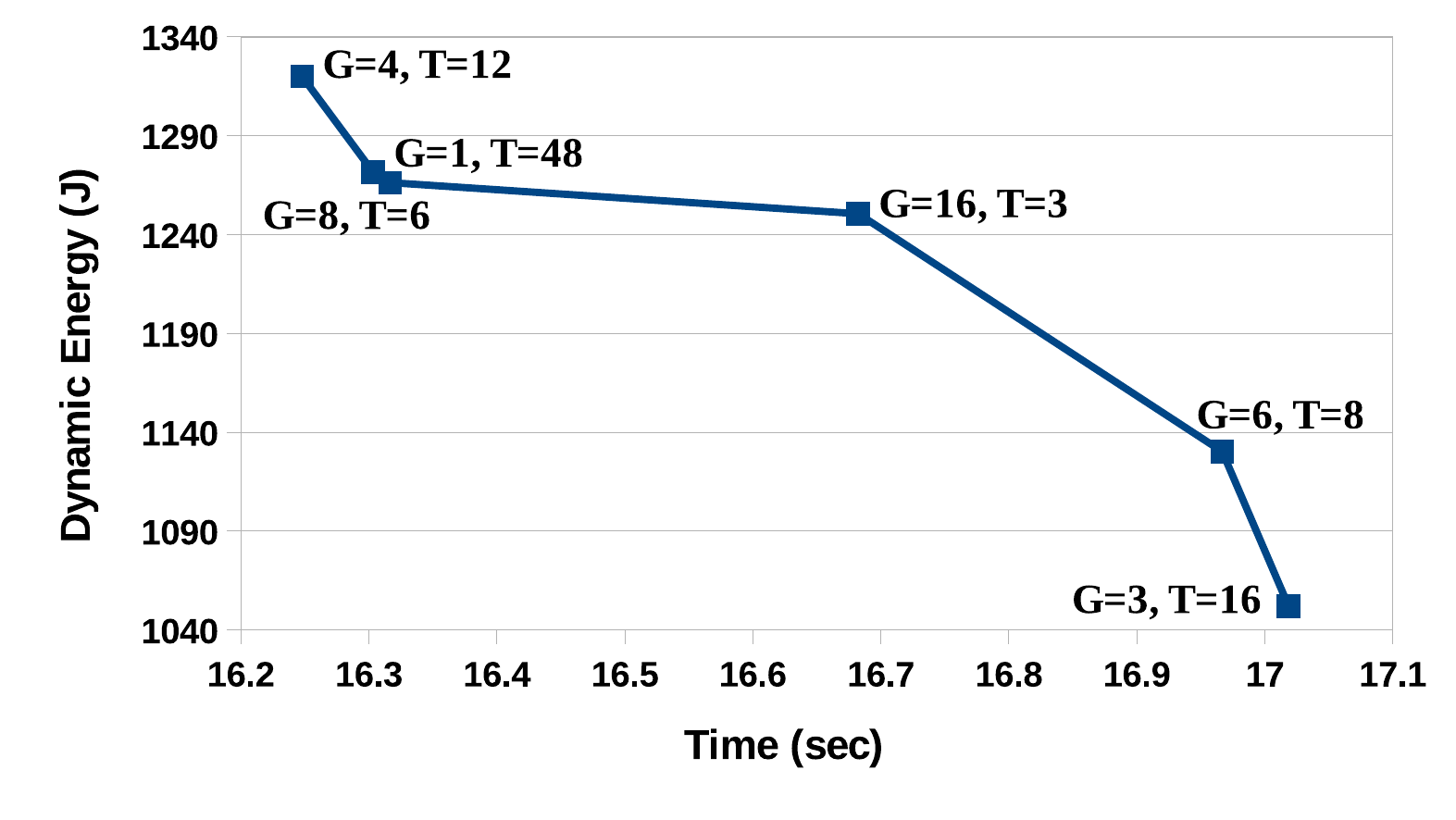}
\label{fig:energy_17408_openblasdgemm}}
\hfill
\caption{(a). Pareto-optimal front of Intel MKL DGEMM PMMTG application on S2 for workload size, N=17408. (b). Pareto-optimal front of OpenBLAS DGEMM PMMTG application on S2 for workload size, N=17408.}
\end{figure}

Figure \ref{fig:energy_17408_intel} shows the globally Pareto-optimal front for PMMTG employing Intel MKL DGEMM on S2 for workload size, N=17408. Optimizing for dynamic energy consumption alone degrades performance by 5.5\%, and optimizing for performance alone increases dynamic energy consumption by 50.7\%. The average and maximum sizes of the Pareto-optimal fronts are (1.8, 4).

Figure \ref{fig:energy_17408_openblasdgemm} shows the globally Pareto-optimal front for PMMTG based on OpenBLAS DGEMM on S2 for workload size, N=17408. There are six globally Pareto-optimal solutions. Optimizing for dynamic energy consumption alone degrades performance by around 5\%, and optimizing for performance alone increases dynamic energy consumption by 20\%. The average and maximum sizes of the Pareto-optimal fronts are 2.4 and 5.

The execution time of building the four dimensional discrete graph with performance and dynamic energy as two objectives and the two decision variables can be cost-prohibitive for its employment in dynamic schedulers and self-adaptable data-parallel applications. We will explore approaches to reduce this time in our future work.

\subsection{Analysis Using Performance and Dynamic Energy Models} \label{sec:pmcs}

In this section, we propose a qualitative dynamic energy model employing performance monitoring counters (PMCs) as parameters. The model reveals the cause behind the energy nonproportionality in modern multicore CPUs. The model along with the execution time of the application is used to analyze the Pareto-optimal front determined by our solution method on a dual-socket multicore platform.

PMCs are special-purpose registers provided in modern microprocessors to store the counts of software and hardware activities. The acronym PMCs is used to refer to software events, which are pure kernel-level counters such as \emph{page-faults}, \emph{context-switches}, etc. as well as micro-architectural events originating from the processor and its performance monitoring unit called the hardware events such as \emph{cache-misses}, \emph{branch-instructions}, etc. Software energy predictive models based on PMCs is one of the leading methods of measurement of energy consumption of an application \cite{FahdA-en12112204}.

The experimental platform S2 and the application OpenBLAS-DGEMM is employed for the analysis. Likwid tool \cite{treibig2010likwid} is used to obtain the PMCs. On this platform, it offers 164 PMCs, which are divided into 28 groups (L2CACHE, L3CACHE, NUMA, etc.). The groups are listed in the supplemental. All the PMCs for each workload size executed using different application configurations, (\#threadgroups (g), \#threads\_per\_group (t)) are collected. Each PMC value is the average for all the 24 physical cores. We analyzed the data to identify the major performance groups, which are highly correlated with the dynamic energy consumption. The highest correlation is contained in the data provided by TLB\_DATA performance group. This group provides data activity, such as load miss rate, store miss rate and walk page duration, in L1 data translation lookaside buffer (dTLB), a small specialized cache of recent page address translations. If a dTLB miss occurs, the OS goes through the page tables. If there is a miss from the page walk, a page fault occurs resulting in the OS retrieving the corresponding page from memory. The duration of the page walk has the highest positive correlation with dynamic energy consumption based on our experiments.

\begin{figure}[!t]
\centering
\subfloat[][]{
\includegraphics[width=1\linewidth]{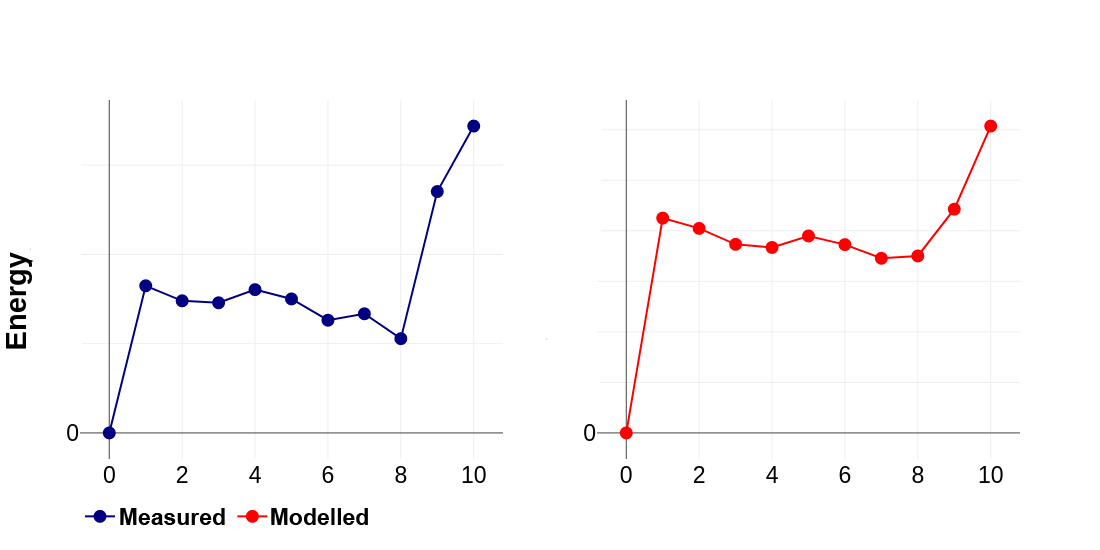}
\label{fig:16384_predicted}}
\hfill
\subfloat[][]{
\includegraphics[width=1\linewidth]{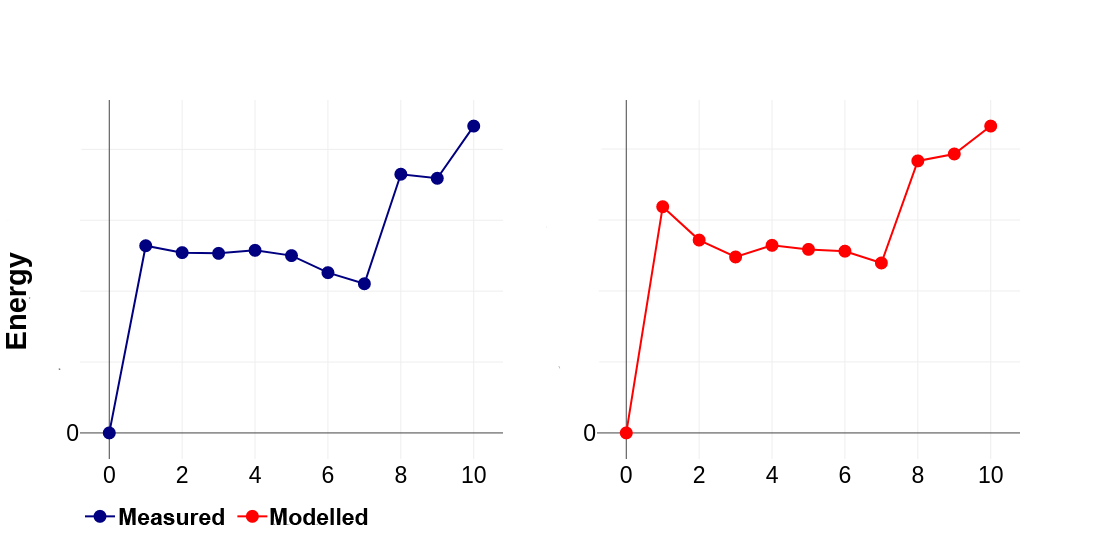}
\label{fig:17408_predicted}}
\hfill
\caption{(a). Measured (left) and predicted (right) dynamic energy consumption of OpenBLAS DGEMM on S2 for workload size, N=16384. (b). Measured (left) and predicted (right) dynamic energy consumption of OpenBLAS DGEMM on S2 for workload size, N=17408.}
\end{figure}

\begin{table*}[!t]
	\caption{L1 dTLB PMC data for size 16384}
	\label{table:PerfData_16384}
	\centering
	\begin{tabular}{ |l|l|l|l|l| }
		\hline
		\textbf{Combination (g, t)} & \textbf{Dynamic Energy (J)} & \textbf{Time (sec)} & \textbf{L1 dTLB load miss duration (Cyc)} & \textbf{L1 dTLB store miss duration (Cyc)} \\ \hline
		\textcolor{blue}{(1,48)} & \textcolor{blue}{824.2743} & \textcolor{blue}{14.112} & \textcolor{blue}{108.373} & \textcolor{blue}{124.326} \\ \hline 
		\textcolor{blue}{(4,12)} & \textcolor{blue}{740.0211} & \textcolor{blue}{14.177} & \textcolor{blue}{113.515} & \textcolor{blue}{105.363} \\ \hline
		\textcolor{blue}{(8,6)} & \textcolor{blue}{729.1005} & \textcolor{blue}{14.244} & \textcolor{blue}{104.564} & \textcolor{blue}{89.3753} \\ \hline
		(2,24) & 802.6687 & 14.314 & 105.328 & 82.5185 \\ \hline
		(16,3) & 750.6159 & 14.615 & 100.924 & 90.2733 \\ \hline
	    \textcolor{blue}{(3,16)} & \textcolor{blue}{631.3098} & \textcolor{blue}{14.772} & \textcolor{blue}{97.9180} & \textcolor{blue}{76.1889} \\ \hline
		 (6,8) & 667.4856 & 14.818 & 96.8957 & 58.0210 \\ \hline
		\textcolor{blue}{(12,4)} & \textcolor{blue}{528.0411} & \textcolor{blue}{15.057} & \textcolor{blue}{97.0492} & \textcolor{blue}{52.8966} \\ \hline
		(24,2) & 1352.141 & 15.875 & 100.106 & 82.7514 \\ \hline
		(48,1) & 1719.012 & 18.685 & 111.902 & 85.9282 \\ \hline
	\end{tabular}
\end{table*}

\begin{table*}[!t]
	\caption{L1 dTLB PMC data for size 17408}
	\label{table:PerfData_17408}
	\centering
	\begin{tabular}{ |l|l|l|l|l| }
		\hline
		\textbf{Combination (g, t)} & \textbf{Dynamic Energy (J)} & \textbf{Time (sec)} & \textbf{L1 dTLB load miss duration (Cyc)} & \textbf{L1 dTLB store miss duration (Cyc)} \\ \hline
		\textcolor{blue}{(4,12)} & \textcolor{blue}{1320.0702} & \textcolor{blue}{16.2478} & \textcolor{blue}{105.961} & \textcolor{blue}{122.191} \\ \hline
		\textcolor{blue}{(1,48)} & \textcolor{blue}{1271.5506} & \textcolor{blue}{16.3034} & \textcolor{blue}{99.5398} & \textcolor{blue}{63.7090} \\ \hline
		\textcolor{blue}{(8,6)} & \textcolor{blue}{1266.3294} & \textcolor{blue}{16.3166} & \textcolor{blue}{95.7896} & \textcolor{blue}{58.9096} \\ \hline
		(2,24) & 1287.6882 & 16.4498 & 98.2180 & 74.6859 \\ \hline
		\textcolor{blue}{(16,3)} & \textcolor{blue}{1250.5616} & \textcolor{blue}{16.6824} & \textcolor{blue}{95.2988} & \textcolor{blue}{58.3551} \\ \hline
	    \textcolor{blue}{(6,8)} & \textcolor{blue}{1130.2412 } & \textcolor{blue}{16.9668 }& \textcolor{blue}{93.4336 }& \textcolor{blue}{47.9097 }\\ \hline
		\textcolor{blue}{(3,16}) &\textcolor{blue}{ 1052.0283}  &\textcolor{blue}{ 17.0187} &\textcolor{blue}{ 90.5275} &\textcolor{blue}{ 45.7483} \\ \hline
		(24,2) & 1824.5795 & 18.0755 & 106.804 & 55.5686 \\ \hline
		(12,4) & 1795.7680 & 20.5520 & 93.6595 & 46.5541 \\ \hline
		(48,1) & 2164.1212 & 20.9868 & 96.6999 & 71.4943 \\ \hline
	\end{tabular}
\end{table*}

Non-negative multivariate regression is employed to construct our model of dynamic energy consumption based on the PMC data from dTLB. The model is shown below:

\begin{equation}
\label{equ:linear}
	E_{dynamic} = \beta_1 \times T + \beta_2 \times L + \beta_3 \times S
\end{equation}

\noindent where $\beta_1$ is the average CPU utilization, $\beta_2$ and $\beta_3$ are the regression coefficients for the PMC data. $T$ is the execution time of the application, $L$ is the time of page walk caused by load miss and $S$ is the time of page walk caused by store miss in dTLB. The coefficients of the model (\{$\beta_1,\beta_2,\beta_3$\}) are forced to be non-negative to avoid erroneous cases where large values for them gives rise to negative dynamic energy consumption prediction violating the fundamental energy conservation law of computing.

To test this model, we use two workload sizes 16384 and 17408. The PMC data that is obtained for these sizes and that is used to train the model is shown in the tables \ref{table:PerfData_16384} and \ref{table:PerfData_17408}. The rows of the tables are sorted in increasing order of time. The blue colour in the tables shows the rows that are in the Pareto-optimal front. The time of page walk (last two columns, 4 and 5) is measured in cycles. As can be seen from the tables, the dynamic energy decreases as the number of cycles decreases.There is however a trade-off between the execution time of application and the page walk time. For a Pareto-optimal solution, a long execution time corresponds to smaller number of load and store cycles and thereby less dynamic energy consumption.

Two dynamic energy models for the workload sizes 16384 (Table \ref{table:PerfData_16384}) and 17408 (Table \ref{table:PerfData_17408}) were constructed. The coefficients for the workload size 16384 are \{$\beta_1 = 253.680, \beta_2 = 39.536, \beta_3 = 13.647$\}. The coefficients for the workload size 17408 are \{$\beta_1 = 137.953, \beta_2 = 12.564, \beta_3 = 3.835$\}. We then predict the dynamic energy consumption using the model and compare with the dynamic energy measured using HCLWattsUp. The Figures \ref{fig:16384_predicted} and \ref{fig:17408_predicted} illustrate the comparison. The $x$ axis represents the number of a row in the Tables \ref{table:PerfData_16384}, \ref{table:PerfData_17408}. The modeled dynamic energy demonstrates the same trend as the measured dynamic energy using HCLWattsUp.

TLB activity has been the focus of research in \cite{kadayif2007reducing,karakostas2016energy,karakostas2015redundant} where the authors state that the address translation using the TLB consumes as much as 16\% of the chip power on some processors. The authors propose different strategies to improve the reuse of TLB caches. Our solution method employing threadgroups (or grouping using multithreaded kernels) allows to fill the page tables more evenly and reduce the duration of page walk resulting in less dynamic energy consumption.

To summarize, our proposed dynamic model based on parameters reflecting TLB activity (the duration of page walk) shows that the energy nonproportionality on our experimental platforms for the data-parallel applications is due to the activity of the data translation lookaside buffer (dTLB), which is disproportionately energy expensive. This finding may encourage the chip design architects to investigate and remove the nonproportionality in these platforms. There may be other causes behind the lack of energy proportionality as the range of applications and platforms is broadened that we would explore in our future research.

\section{Conclusion} \label{sec:conclusions}

Energy proportionality is the key design goal followed by architects of modern multicore CPUs. One of its implications is that optimization of an application for performance will also optimize it for energy. However, due to the inherent complexities of resource contention for shared on-chip resources, NUMA, and dynamic power management in multicore CPUs, state-of-the-art application-level optimization methods for performance and energy \cite{LastovetskyReddy2017,manumachu2018bi,manumachu2018bicpe,Hamid2019}, demonstrate that the functional relationships between performance and workload size and between dynamic energy and workload size for real-life data-parallel applications have complex (non-linear) properties and show that workload distribution has become an important decision variable. 

This motivated us to explore in-depth the influence of three-dimensional decision variable space on bi-objective optimization of applications for performance and energy on multicore CPUs. The three decision variables are: a). The number of identical multithreaded kernels (threadgroups) involved in the parallel execution of an application; b). The number of threads in each threadgroup; and c). The workload distribution between the threadgroups. We focused exclusively on the first two decision variables in this work. 

By experimenting with these decision variables, we discovered that energy proportionality does not hold true for modern multicore CPUs. Based on this finding, we proposed the first application-level optimization method for bi-objective optimization of multithreaded data-parallel applications for performance and energy on a single multicore CPU. The method uses two decision variables, the number of identical multithreaded kernels (threadgroups) and the number of threads in each threadgroup. A given workload is partitioned equally between the threadgroups.

We demonstrated our method using four highly optimized multithreaded data-parallel applications, 2D fast Fourier transform based on FFTW and Intel MKL, and dense matrix-matrix multiplication written using Openblas DGEMM and Intel MKL, on four modern multicore CPUs one of which is a single socket multicore CPU and the other three dual-socket with increasing number of physical cores per socket. We showed in particular that optimizing for performance alone results in significant increase in dynamic energy consumption whereas optimizing for dynamic energy alone results in considerable performance degradation and that our method determined good number of globally Pareto-optimal solutions.

Finally, we proposed a qualitative dynamic energy model employing performance monitoring counters (PMCs) as parameters, which we used to explain the Pareto-optimal solutions determined for modern multicore CPUs. The model showed that the energy nonproportionality on our experimental platforms for the two data-parallel applications is caused by disproportionately high energy consumption by the data translation lookaside buffer (dTLB) activity.

\ifCLASSOPTIONcompsoc
\section*{Acknowledgments}
\else
\section*{Acknowledgment}
\fi

This publication has emanated from research conducted with the financial support of Science Foundation Ireland (SFI) under Grant Number 14/IA/2474. We thank Roman Wyrzykowski and Lukasz Szustak for allowing us to use their Intel servers, HCLServer03 and HCLServer04.

\section{Supplementary Material}

\subsection{Rationale Behind Using Dynamic Energy Consumption Instead of Total Energy Consumption} \label{denergy-rationale}

There are two types of energy consumptions, static energy, and dynamic energy. We define the static energy consumption as the energy consumption of the platform without the given application execution. Dynamic energy consumption is calculated by subtracting this static energy consumption from the total energy consumption of the platform during the given application execution.  The static energy consumption is calculated by multiplying the idle power of the platform (without application execution) with the execution time of the application. That is, if $P_S$ is the static power consumption of the platform, $E_T$ is the total energy consumption of the platform during the execution of an application, which takes $T_E$ seconds, then the dynamic energy $E_D$ can be calculated as,

\begin{equation} \label{eq:DynamicEnergy}
\begin{aligned}
E_D = E_T - (P_S \times T_E)
\end{aligned}
\end{equation}

We consider only the dynamic energy consumption in our work for reasons below:

\begin{enumerate}
	\item Static energy consumption is a constant (or a inherent property) of a platform that can not be optimized. It does not depend on the application configuration.
	\item Although static energy consumption is a major concern in embedded systems, it is becoming less compared to the dynamic energy consumption due to advancements in hardware architecture design in HPC systems.
	\item We target applications and platforms where dynamic energy consumption is the dominating energy dissipator.
	\item Finally, we believe its inclusion can underestimate the true worth of an optimization technique that minimizes the dynamic energy consumption. We elucidate using two examples from published results.
	\begin{itemize}
		\item In our first example, consider a model that reports predicted and measured total energy consumption of a system to be 16500J and 18000J. It would report the prediction error to be 8.3\%. If it is known that the static energy consumption of the system is 9000J, then the actual prediction error (based on dynamic energy consumptions only) would be 16.6\% instead.
		\item In our second example, consider two different energy prediction models ($M_A$ and $M_B$) with same prediction errors of 5\% for an application execution on two different machines ($A$ and $B$) with same total energy consumption of 10000J. One would consider both the models to be equally accurate. But supposing it is known that the dynamic energy proportions for the machines are 30\% and 60\%. Now, the true prediction errors (using dynamic energy consumptions only) for the models would be 16.6\% and 8.3\%. Therefore, the second model $M_B$ should be considered more accurate than the first.
	\end{itemize}
\end{enumerate}

\subsection{Shared Memory Implementations of PMMTG Algorithms}

The shared memory implementations of PMMTG algorithms using Intel MKL and OpenBLAS are described here. The inputs to an implementation are: a). Matrices A, B, and C of sizes $N \times N$; b). Constants $\alpha$ and $\beta$; c) The number of abstract processors (groups), $\{P_1,\cdots,P_p\}$; d). The number of threads in each abstract processor (group) represented by $t$. The output matrix, C, contains the matrix product. Each abstract processor is a group of $t$ threads.

The implementations using Intel MKL differ from those using OpenBLAS. In Intel MKL, the matrix-matrix computation specific to a partition is computed using an OpenMP parallel region with $t$ threads whereas the same is computed in OpenBLAS using a pthread.

\subsubsection{Intel MKL implementation of PMMTG-V}

Figure \ref{fig:intel-mkl-pmmtg-v} shows the implementation of PMMTG-V using Intel MKL.

\begin{figure}[!t]
\lstset{language=C++}
\begin{lstlisting}
int row;
#pragma omp parallel for num_threads(p*t)
for (row = 0; row < N; row++) {
    memcpy(&B1[row*(N/p)], &B[row*N],
           (N/p)*sizeof(double));
    ...
    memcpy(&Bp[row*(N/p)], &B[(p-1)*(N/p)+row*N], 
           (N/p)*sizeof(double));
    memcpy(&C1[row*(N/p)], &C[row*N], 
           (N/p)*sizeof(double));
    ...
    memcpy(&Cp[row*(N/p)], &C[(p-1)*(N/p)+row*N], 
           (N/p)*sizeof(double));
}

#pragma omp parallel sections num_threads(p*t)
{
	#pragma omp section // processor 1
	{
	    mkl_set_num_threads_local(t);
	    cblas_dgemm(CblasRowMajor, CblasNoTrans, 
	      CblasNoTrans, N, N/p, N, alpha, A, N, 
	      B1, N/p, beta, C1, N/p);
	}
	...
	#pragma omp section // processor p
	{
	    mkl_set_num_threads_local(t);
	    cblas_dgemm(CblasRowMajor, CblasNoTrans, 
	         CblasNoTrans, N, N/p, N, alpha, A, N, 
	         Bp, N/p, beta, Cp, N/p);
	}
}

#pragma omp parallel for num_threads(p*t)
for (row = 0; row < N; row++)
{
    memcpy(&B[row*N], &B1[row*(N/p)], 
      (N/p)*sizeof(double));
    ...
    memcpy(&B[(p-1)*(N/p)+row*N], &Bp[row*(N/p)], 
           (N/p)*sizeof(double));
    memcpy(&C[row*N], &C1[row*(N/p)], 
           (N/p)*sizeof(double));
    ...
    memcpy(&C[(p-1)*(N/p)+row*N], &Cp[row*(N/p)], 
           (N/p)*sizeof(double));
}
\end{lstlisting}
\caption{Intel MKL implementation of PMMTG-V employing $p$ abstract processors of $t$ threads each.}
\label{fig:intel-mkl-pmmtg-v}
\end{figure}

\subsubsection{OpenBLAS implementation of PMMTG-V}

Figure \ref{fig:openblas-pmmtg-v} shows the implementation of PMMTG-V using OpenBLAS.

\begin{figure}[!t]
\lstset{language=C++}
\begin{lstlisting}
void *dgemm(void *input) {
    int i = *(int*)input;
    openblas_set_num_threads(t);
    goto_set_num_threads(t);
    omp_set_num_threads(t);
    if (i == 1)
    {
  		cblas_dgemm(CblasRowMajor, CblasNoTrans, 
  		    CblasNoTrans, N, N/p, N, alpha, A, N, 
  		    B1, N/p, beta, C1, N/p);
   	}
   	...
    if (i == p)
   	{
   		cblas_dgemm(CblasRowMajor, CblasNoTrans, 
   		    CblasNoTrans, N, N/p, N, alpha, A, N, 
   		    Bp, N/p, beta, Cp, N/p);
    }
}

int main() {
    int row;
#pragma omp parallel for num_threads(p*t)
    for (row = 0; row < N; row++) {
    	memcpy(&B1[row*(N/p)], &B[row*N], 
    	       (N/p)*sizeof(double));
    	...
    	memcpy(&Bp[row*(N/p)], &B[(p-1)*(N/p)+row*N], 
    	       (N/p)*sizeof(double));
    	memcpy(&C1[row*(N/p)], &C[row*N], 
    	       (N/p)*sizeof(double));
    	...
    	memcpy(&Cp[row*(N/p)], &C[(p-1)*(N/p)+row*N], 
    	       (N/p)*sizeof(double));
    }
    
    pthread_t t1, ..., tp;
    int i1 = 1, ..., ip = p;
    pthread_create(&t1, NULL, dgemm, &i1);
    ...
    pthread_create(&tp, NULL, dgemm, &ip);
    pthread_join(tp, NULL);
    ...
    pthread_join(t1, NULL);
    
#pragma omp parallel for num_threads(p*t)
    for (row = 0; row < N; row++)
    {
    	memcpy(&B[row*N], &B1[row*(N/p)], 
    	       (N/p)*sizeof(double));
    	...
    	memcpy(&B[(p-1)*(N/p)+row*N], &Bp[row*(N/p)], 
    	       (N/p)*sizeof(double));
    	memcpy(&C[row*N], &C1[row*(N/p)], 
    	       (N/p)*sizeof(double));
    	...
    	memcpy(&C[(p-1)*(N/p)+row*N], &Cp[row*(N/p)], 
    	       (N/p)*sizeof(double));
    }
}
\end{lstlisting}
\caption{OpenBLAS implementation of PMMTG-V employing $p$ abstract processors of $t$ threads each.}
\label{fig:openblas-pmmtg-v}
\end{figure}

\subsubsection{Intel MKL implementation of PMMTG-S}

Figure \ref{fig:intel-mkl-pmmtg-s} shows the implementation of PMMTG-S using Intel MKL.

\begin{figure}[!t]
\lstset{language=C++}
\begin{lstlisting}
#pragma omp parallel for num_threads(4*t)
for (row = 0; row < (N/2); row++) {
    memcpy(&A11[row*(N/2)], &A[row*N], 
           (N/2)*sizeof(double));
    memcpy(&A22[row*(N/2)], &A[N*(N/2)+(N/2)+row*N], 
           (N/2)*sizeof(double));
    ...
    memcpy(&B11[row*(N/2)], &B[row*N], 
           (N/2)*sizeof(double));
    memcpy(&B22[row*(N/2)], &B[N*(N/2)+(N/2)+row*N], 
           (N/2)*sizeof(double));
    ...
    memcpy(&C11[row*(N/2)], &C[row*N], 
           (N/2)*sizeof(double));
    memcpy(&C22[row*(N/2)], &C[N*(N/2)+(N/2)+row*N], 
           (N/2)*sizeof(double));
   	...
}

#pragma omp parallel sections num_threads(4)
{
   	#pragma omp section // processor 1
   	{
        mkl_set_num_threads_local(t);
        cblas_dgemm(CblasRowMajor, CblasNoTrans, 
           CblasNoTrans, N/2, N/2, N/2, alpha, A11, N/2, 
           B11, N/2, beta0, C11, N/2);
        cblas_dgemm(CblasRowMajor, CblasNoTrans, 
           CblasNoTrans, N/2, N/2, N/2, alpha, A12, N/2, 
           B21, N/2, beta1, C11, N/2);
   	}
   	...
   	#pragma omp section // processor 4
   	{
        mkl_set_num_threads_local(t);    		
        cblas_dgemm(CblasRowMajor, CblasNoTrans, 
           CblasNoTrans, N/2, N/2, N/2, alpha, A21, N/2, 
           B12, N/2, beta0, C22, N/2);
        cblas_dgemm(CblasRowMajor, CblasNoTrans, 
           CblasNoTrans, N/2, N/2, N/2, alpha, A22, N/2, 
           B22, N/2, beta1, C22, N/2);
   	}
}

#pragma omp parallel for num_threads(4*t)
for (row = 0; row < (N/2); row++) {
   memcpy(&A[row*N], &A11[row*(N/2)], 
          (N/2)*sizeof(double));
   memcpy(&A[N*(N/2)+(N/2)+row*N], &A22[row*(N/2)], 
          (N/2)*sizeof(double));
   ...
   memcpy(&B[row*N], &B11[row*(N/2)], 
          (N/2)*sizeof(double));
   memcpy(&B[N*(N/2)+(N/2)+row*N], &B22[row*(N/2)], 
          (N/2)*sizeof(double));
   ...
   memcpy(&C[row*N], &C11[row*(N/2)], 
          (N/2)*sizeof(double));
   memcpy(&C[N*(N/2)+(N/2)+row*N], &C22[row*(N/2)], 
          (N/2)*sizeof(double));
   ...
}
\end{lstlisting}
\caption{Intel MKL implementation of PMMTG-S employing 4 abstract processors of $t$ threads each and arranged in a 2 $\times$ 2 grid.}
\label{fig:intel-mkl-pmmtg-s}
\end{figure}

\subsubsection{OpenBLAS implementation of PMMTG-S}

Figure \ref{fig:openblas-pmmtg-s} shows the implementation of PMMTG-S using OpenBLAS.

\begin{figure}[!t]
\lstset{language=C++}
\begin{lstlisting}
void *dgemm(void *input){
	int i = *(int*)input;
	openblas_set_num_threads(t);
	goto_set_num_threads(t);
	omp_set_num_threads(t);
	if (i == 1){
       cblas_dgemm(CblasRowMajor, CblasNoTrans, 
         CblasNoTrans, N/2, N/2, N/2, alpha, A11, N/2, 
         B11, N/2, beta0, C11, N/2);
       cblas_dgemm(CblasRowMajor, CblasNoTrans, 
         CblasNoTrans, N/2, N/2, N/2, alpha, A12, N/2, 
         B21, N/2, beta1, C11, N/2);
	}
	...
	if (i == 4){
       cblas_dgemm(CblasRowMajor, CblasNoTrans, 
         CblasNoTrans, N/2, N/2, N/2, alpha, A21, N/2, 
         B12, N/2, beta0, C22, N/2);
       cblas_dgemm(CblasRowMajor, CblasNoTrans, 
         CblasNoTrans, N/2, N/2, N/2, alpha, A22, N/2, 
         B22, N/2, beta1, C22, N/2);	
    }
}
int main(){
	int row;
#pragma omp parallel for num_threads(4*t)
	for (row = 0; row < (N/2); row++) {
        memcpy(&A11[row*(N/2)], &A[row*N], 
               (N/2)*sizeof(double));
        memcpy(&A22[row*(N/2)], &A[N*(N/2)+(N/2)+row*N], 
               (N/2)*sizeof(double));
		...
        memcpy(&B11[row*(N/2)], &B[row*N], 
               (N/2)*sizeof(double));
        memcpy(&B22[row*(N/2)], &B[N*(N/2)+(N/2)+row*N], 
               (N/2)*sizeof(double));
		...
        memcpy(&C11[row*(N/2)], &C[row*N], 
               (N/2)*sizeof(double));
        memcpy(&C22[row*(N/2)], &C[N*(N/2)+(N/2)+row*N], 
               (N/2)*sizeof(double));
		...
	}
	
	pthread_t t1, ..., t4;
	int i1 = 1, ..., i4 = 4;
	pthread_create(&t1, NULL, dgemm, &i1);
	...
	pthread_create(&t4, NULL, dgemm, &i4);
	pthread_join(t4, NULL);
	...
	pthread_join(t1, NULL);
	
#pragma omp parallel for num_threads(4*t)
	for (row = 0; row < (N/2); row++) {
        memcpy(&A[row*N], &A11[row*(N/2)], 
               (N/2)*sizeof(double));
        memcpy(&A[N*(N/2)+(N/2)+row*N], &A22[row*(N/2)], 
               (N/2)*sizeof(double));
        ...
        memcpy(&B[row*N], &B11[row*(N/2)], 
               (N/2)*sizeof(double));
        memcpy(&B[N*(N/2)+(N/2)+row*N], &B22[row*(N/2)], 
               (N/2)*sizeof(double));
        ...
        memcpy(&C[row*N], &C11[row*(N/2)], 
               (N/2)*sizeof(double));
        memcpy(&C[N*(N/2)+(N/2)+row*N], &C22[row*(N/2)], 
               (N/2)*sizeof(double));
        ...
	}
}
\end{lstlisting}
\caption{OpenBLAS implementation of PMMTG-S employing 4 abstract processors of $t$ threads each and arranged in a 2 $\times$ 2 grid.}
\label{fig:openblas-pmmtg-s}
\end{figure}

\subsubsection{Intel MKL implementation of PMMTG-H}

Figure \ref{fig:intel-mkl-pmmtg-h} shows the implementation of PMMTG-H using Intel MKL.

\begin{figure}[!t]
\lstset{language=C++}
\begin{lstlisting}
int row;
#pragma omp parallel for num_threads(p*t)
for (row = 0; row < N/p; row++) {
	memcpy(&A1[row*N], &A[row*N], 
           N*sizeof(double));
	...
	memcpy(&Ap[row*N], &A[(p-1)*N*(N/p)+row*N], 
           N*sizeof(double));
	memcpy(&C1[row*N], &C[row*N], 
           N*sizeof(double));
	...
	memcpy(&Cp[row*N], &C[(p-1)*N*(N/p)+row*N], 
           N*sizeof(double));
}

#pragma omp parallel sections num_threads(p*t)
{
	#pragma omp section // processor 1
	{
		mkl_set_num_threads_local(t);
		cblas_dgemm(CblasRowMajor, CblasNoTrans, 
		  CblasNoTrans, N/p, N, N, alpha, A1, N, 
		  B, N, beta, C1, N);
	}
	...
	#pragma omp section // processor p
	{
		mkl_set_num_threads_local(t);
		cblas_dgemm(CblasRowMajor, CblasNoTrans, 
		  CblasNoTrans, N/p, N, N, alpha, Ap, N, 
		  B, N, beta, Cp, N);
	}
}

#pragma omp parallel for num_threads(p*t)
for (row = 0; row < N/p; row++)
{
	memcpy(&A[row*N], &A1[row*N], 
	       N*sizeof(double));
	...
	memcpy(&A[(p-1)*N*(N/p)+row*N], &Ap[row*N], 
	       N*sizeof(double));
	memcpy(&C[row*N], &C1[row*N], 
	       N*sizeof(double));
	...
	memcpy(&C[(p-1)*N*(N/p)+row*N], &Cp[row*N], 
	       N*sizeof(double));
}
\end{lstlisting}
\caption{Intel MKL implementation of PMMTG-H employing $p$ abstract processors of $t$ threads each.}
\label{fig:intel-mkl-pmmtg-h}
\end{figure}

\subsection{Shared Memory Implementations of PFFT Algorithms} \label{pfft-pseudocodes}

The inputs to an implementation are: a). Signal matrix $\mathcal{M}$ of size $N \times N$; b). The number of abstract processors (groups) $p$, $\{P_1,\cdots,P_p\}$; c). The number of threads in each abstract processor (group) represented by $t$. The output is the transformed signal matrix $\mathcal{M}$ (considering that we are performing in-place FFT). Each abstract processor is a group of $t$ threads.

The implementations using Intel MKL differ from those using FFTW. In FFTW, only plan execution (fftw\_plan\_many\_dft) and plan destruction (fftw\_destroy\_plan) are thread-safe and called be called in an OpenMP parallel region.

\subsubsection{Intel MKL implementation of PFFTTG-H}

Figure \ref{fig:intel-mkl-pffttg-h} shows the implementation of PFFTTG-H using Intel MKL.

\begin{figure}[!t]
\lstset{language=C++}
\begin{lstlisting}

void fftw1d(const int sign, const int m,
    const int n, fftw_complex* X,fftw_complex* Y)
{
    int rank = 1, howmany = m;
    int s[] = {n};
    int idist = n, odist = n;
    int istride = 1, ostride = 1;
    int *inembed = s, *onembed = s;
    fftw_plan my_plan = fftw_plan_many_dft(
                            rank, s, howmany,
                            X, inembed, istride, idist,
                            Y, onembed, ostride, odist,
                            sign, FFTW_ESTIMATE);
    fftw_execute(my_plan);
    fftw_destroy_plan(my_plan);
    return;
}

int
fftw2d(const int sign, const int N, const int p,
    const unsigned int nt, const unsigned int blockSize,
    fftw_complex* X)
{
#pragma omp parallel sections num_threads(p)
{
    #pragma omp section
    {
       fftw1d(sign, N/p, N, X, X);
    }
...
    #pragma omp section
    {
       fftw1d(sign, N-(p-1)*(N/p), N, 
              &X[(p-1)*(N/p)*N], &X[(p-1)*(N/p)*N]);
    }
}

    hcl_transpose_block(X, 0, N, N, nt, blockSize);

#pragma omp parallel sections num_threads(p)
{
    #pragma omp section
    {
       fftw1d(sign, N/p, N, X, X);
    }
...
    #pragma omp section
    {
       fftw1d(sign, N-(p-1)*(N/p), N, 
              &X[(p-1)*(N/p)*N], &X[(p-1)*(N/p)*N]);
    }
}

    hcl_transpose_block(X, 0, N, N, nt, blockSize);
}

\end{lstlisting}
\caption{Intel MKL implementation of PFFTTG-H employing $p$ abstract processors of $t$ threads each.}
\label{fig:intel-mkl-pffttg-h}
\end{figure}

\subsection{Transpose Routine Invoked in PFFT Algorithms} \label{transpose}

The routine, hcl\_transpose\_block, shown in the Figure \ref{fig:transpose} performs in-place transpose of a complex 2D square matrix of size $n \times n$. We use a block size of 64 in our experiments as it is found to be optimal.

\begin{figure}[!t]
\lstset{language=C++}
\begin{lstlisting}
void hcl_transpose_scalar_block(fftw_complex* X1, 
    fftw_complex* X2, const int i, const int j, 
    const int N, const int block_size)
{
    int p, q;
    for (p = 0; p < min(N-i,block_size); p++) {
        for (q = 0; q < min(N-j,block_size); q++) {
           int index1 = i*N+j + p*N+q;
           int index2 = j*N+i + q*N+p;
           
           if (index1 >= index2)
              continue;
        
           double tmpr = X1[p*N+q][0];
           double tmpi = X1[p*N+q][1];
           X1[p*N+q][0] = X2[q*N+p][0];
           X1[p*N+q][1] = X2[q*N+p][1];
           X2[q*N+p][0] = tmpr;
           X2[q*N+p][1] = tmpi;
        }
    }
}

void hcl_transpose_block(fftw_complex* X, const int start, 
    const int end, const int n, 
    const unsigned int nt, const int block_size)
{
    int i, j;
#pragma omp parallel for shared(X) private(i, j) num_threads(nt)
    for (i = 0; i < end; i += block_size) {
        for (j = 0; j < end; j += block_size) {
            hcl_transpose_scalar_block(
                &X[start + i*N + j],
                &X[start + j*N + i], 
                i, j, N, block_size);
        }
    }
}

\end{lstlisting}
\caption{Transpose of square matrix of size $n \times n$ using blocking.}
\label{fig:transpose}
\end{figure}

\vfill
	
\subsection{Application Programming Interface (API) for Measurements Using External Power Meter Interfaces (HCLWattsUp)} \label{appendix:HCLWattsUp API}

HCLServer01, HCLServer02 and HCLServer03 have a dedicated power meter installed between their input power sockets and wall A/C outlets. The power meter captures the total power consumption of the node. It has a data cable connected to the USB port of the node. A perl script collects the data from the power meter using the serial USB interface. The execution of this script is non-intrusive and consumes insignifcant power.

We use \textit{HCLWattsUp} API function, which gathers the readings from the power meters to determine the average power and energy consumption during the execution of an application on a given platform. \textit{HCLWattsUp} API can provide following four types of measures during the execution of an application:
\begin{itemize}
	\item \textit{TIME}---The execution time (seconds).
	\item \textit{DPOWER}---The average dynamic power (watts).
	\item \textit{TENERGY}---The total energy consumption (joules).
	\item \textit{DENERGY}---The dynamic energy consumption (joules).
\end{itemize}

We confirm that the overhead due to the API is very minimal and does not have any noticeable influence on the main measurements. It is important to note that the power meter readings are only processed if the measure is not $\textit{hcl::TIME}$. Therefore, for each measurement, we have two runs. One run for measuring the execution time. And the other for energy consumption. The following example illustrates the use of statistical methods to measure the dynamic energy consumption during the execution of an application.

The API is confined in the \textit{hcl} namespace. Lines 10--12 construct the \textup{Wattsup} object. The inputs to the constructor are the paths to the scripts and their arguments that read the USB serial devices containing the readings of the power meters.

The principal method of \textit{Wattsup} class is \textit{execute}. The inputs to this method are the type of measure, the path to the executable \textit{executablePath}, the arguments to the executable \textit{executableArgs} and the statistical thresholds (\textit{pIn}) The outputs are the achieved statistical confidence \textit{pOut}, the estimators, the sample mean (\textit{sampleMean}) and the standard deviation (\textit{sd}) calculated during the execution of the~executable.
\begin{figure}[!t]
\lstset{language=C++}
\begin{lstlisting}
#include <wattsup.hpp>
int main(int argc, char** argv)
{
    std::string pathsToMeters[2] = {
       "/opt/powertools/bin/wattsup1", 
       "/opt/powertools/bin/wattsup2"};
    std::string argsToMeters[2] = {
       "--interval=1", 
       "--interval=1"};       
    hcl::Wattsup wattsup(
        2, pathsToMeters, argsToMeters
    );
    hcl::Precision pIn = {
        maxRepeats, cl, maxElapsedTime, maxStdError
    };
    hcl::Precision pOut;
    double sampleMean, sd;
    int rc = wattsup.execute(
               hcl::DENERGY, executablePath, 
               executableArgs, &pIn, &pOut,
               &sampleMean, &sd
    );
    if (rc == 0)
       std::cerr << "Precision NOT achieved.\n";
    else
       std::cout << "Precision achieved.\n";
    std::cout << "Max repetitions "
              << pOut.reps_max
              << ", Elasped time "
              << pOut.time_max_rep
              << ", Relative error "
              << pOut-eps-converted-to.pdf
              << ", Mean energy "
              << sampleMean
              << ", Standard Deviation "
              << sd
              << std::endl;
    exit(EXIT_SUCCESS);              
}
\end{lstlisting}
\captionsetup{width=1\linewidth}\caption{Example illustrating the use of HCLWattsUp API for measuring the dynamic energy consumption }
\label{figure A1}
\end{figure}

The \textit{execute} method repeatedly invokes the executable until one of the following conditions is~satisfied:
\begin{itemize}
	\item The maximum number of repetitions specified in $maxRepeats$ is exceeded.
	\item The sample mean is within $maxStdError$ percent of the confidence interval $cl$. The confidence interval of the mean is estimated using Student's t-distribution.
	\item The maximum allowed time $maxElapsedTime$ specified in seconds has elapsed.
\end{itemize}

If any one of the conditions are not satisfied, then a return code of 0 is output suggesting that statistical confidence has not been achieved. If statistical confidence has been achieved, then the number of repetitions performed, time elapsed and the final relative standard error is returned in the output argument $pOut$. At the same time, the sample mean and standard deviation are returned. For our experiments, we use values of (1000, 95\%, 2.5\%, 3600) for the parameters ($maxRepeats,cl,maxStdError,maxElapsedTime$) respectively. Since we use Student's t-distribution for the calculation of the confidence interval of the mean, we confirm specifically that the observations follow normal distribution by plotting the density of the observations using \textit{R} tool.

\subsection{Experimental Methodology to Determine the Sample Mean} \label{exp-methodology}

We followed the methodology described below to make sure the experimental results are reliable:
\begin{itemize}
	\item The server is fully reserved and dedicated to these experiments during their execution. We also made certain that there are no drastic fluctuations in the load due to abnormal events in the server by monitoring its load continuously for a week using the tool \textit{sar}. Insignificant variation in the load was observed during this monitoring period suggesting normal and clean behaviour of the server.
	\item An application during its execution is bound to the physical cores using the \textit{numactl} tool.
	\item To obtain a data point, the application is repeatedly executed until the sample mean lies in the 95\% confidence interval with precision of 0.025 (2.5\%). For this purpose, we use Student's t-test assuming that the individual observations are independent and their population follows the normal distribution. We verify the validity of these assumptions using Pearson's chi-squared test. When we mention a single number such as execution time (seconds) or floating-point performance (in MFLOPs or GFLOPs), we imply the sample mean determined using the Student's t-test.
	
	The function $MeanUsingTtest$, shown in Algorithm \ref{mean-t-test}, determines the sample mean for a data point. For each data point, the function repeatedly executes the application $app$ until one of the following three conditions is satisfied:
	\begin{enumerate}
		\item The maximum number of repetitions ($maxReps$) is exceeded (Line 3).
		\item The sample mean falls in the confidence interval (or the precision of measurement $eps$ is achieved) (Lines 13-15).
		\item The elapsed time of the repetitions of application execution has exceeded the maximum time allowed ($maxT$ in seconds) (Lines 16-18).
	\end{enumerate}
	So, for each data point, the function $MeanUsingTtest$ returns the sample mean $mean$. The function $Measure$ measures the execution time using \emph{gettimeofday} function. 
	\item In our experiments, we set the minimum and maximum number of repetitions, $minReps$ and $maxReps$, to 15 and 100000. The values of $maxT$, $cl$, and $eps$ are 3600, 0.95, and 0.025. If the precision of measurement is not achieved before the completion of maximum number of repeats, we increase the number of repetitions and also the allowed maximum elapsed time. Therefore, we make sure that statistical confidence is achieved for all the data points that we use in our experiments.
\end{itemize}

\begin{algorithm}[!t]
	\caption{Function determining the mean of an experimental run using Student's t-test.}\label{mean-t-test}
	\begin{algorithmic}[1]
		\Procedure{MeanUsingTtest}{$app,minReps,maxReps,$ \par
			$maxT,cl,accuracy,$ \par
			$repsOut,clOut,etimeOut,epsOut,mean$}
		\INPUT
		\Statex The application to execute, $app$
		\Statex The minimum number of repetitions, $minReps \in \mathbb Z_{> 0}$
		\Statex The maximum number of repetitions, $maxReps \in \mathbb Z_{> 0}$
		\Statex The maximum time allowed for the application to run, $maxT \in \mathbb R_{> 0}$
		\Statex The required confidence level, $cl \in \mathbb R_{> 0}$
		\Statex The required accuracy, $eps \in \mathbb R_{> 0}$
		\OUTPUT
		\Statex The number of experimental runs actually made, $repsOut \in \mathbb Z_{> 0}$
		\Statex The confidence level achieved, $clOut \in \mathbb R_{> 0}$
		\Statex The accuracy achieved, $epsOut \in \mathbb R_{> 0}$
		\Statex The elapsed time, $etimeOut \in \mathbb R_{> 0}$
		\Statex The mean, $mean \in \mathbb R_{> 0}$
		\Statex
		\State $reps \gets 0$; $stop \gets 0$; $sum \gets 0$; $etime \gets 0$
		\While{($reps < maxReps$) and ($!stop$)}
		\State $st \gets \Call{measure}{TIME}$
		\State $\Call{Execute}{app}$
		\State $et \gets \Call{measure}{TIME}$
		\State $reps \gets reps + 1$
		\State $etime \gets etime + et-st$
		\State $ObjArray[reps] \gets et-st$
		\State $sum \gets sum + ObjArray[reps]$
		\If{$reps > minReps$}
		\State $clOut$ $\gets$ fabs(gsl\_cdf\_tdist\_Pinv($cl$, $reps-1$)) \par
		\hskip\algorithmicindent\hskip\algorithmicindent\hskip\algorithmicindent
		$\times$ gsl\_stats\_sd($ObjArray$, 1, $reps$) \par
		\hskip\algorithmicindent\hskip\algorithmicindent\hskip\algorithmicindent
		/ sqrt($reps$)
		\If{$clOut \times \frac{reps}{sum} < eps$}
		\State $stop \gets 1$
		\EndIf
		\If{$etime > maxT$}
		\State $stop \gets 1$
		\EndIf
		\EndIf
		\EndWhile
		\State $repsOut \gets reps$; $epsOut \gets clOut \times \frac{reps}{sum}$
		\State $etimeOut \gets etime$; $mean \gets \frac{sum}{reps}$
		\EndProcedure
	\end{algorithmic}
\end{algorithm}

\subsection{List of PMC groups Provided by Likwid} \label{PMC groups}

The list of PMC groups provided by Likwid tool \cite{treibig2010likwid} on HCLServer2 (S2) is shown in the Figure \ref{pmc_groups}.

\begin{table} [!t]
	\caption{Specification of the Intel multicore CPU platform, HCLServer2.}
	\label{table:hclservers}
	\centering
	\begin{tabular}{ |l|l| }
		\hline
		\textbf{Technical Specifications} & \textbf{HCLServer2 (S2) } \\ \hline
		Processor & Intel Haswell E5-2670V3 \\ \hline
		OS & CentOS 7.2.1511 \\ \hline
		Core(s) per socket & 12 \\ \hline
		Socket(s) & 2 \\ \hline
		L1d cache, L1i cache  & 32 KB, 32 KB \\ \hline
		L2 cache, L3 cache & 256 KB, 30976 KB \\ \hline
		Total main memory & 64 GB \\ \hline
		Power meter & WattsUp Pro \\ \hline
	\end{tabular}
\end{table}

\begin{figure}[!t]
\lstset{language=C++}
\begin{lstlisting}
$ likwid-perfctr -a

 Group name	    Description
-----------     ----------------------------------------
     BRANCH	    Branch prediction miss rate/ratio
     CACHES	    Cache bandwidth in MBytes/s
       CBOX	    CBOX related data and metrics
      CLOCK	    Power and Energy consumption
       DATA	    Load to store ratio
     ENERGY	    Power and Energy consumption
FALSE_SHARE	    False sharing
  FLOPS_AVX	    Packed AVX MFLOP/s
         HA	    Main memory bandwidth in MBytes/s seen 
                from Home agent
     ICACHE	    Instruction cache miss rate/ratio
         L2	    L2 cache bandwidth in MBytes/s
    L2CACHE	    L2 cache miss rate/ratio
         L3	    L3 cache bandwidth in MBytes/s
    L3CACHE	    L3 cache miss rate/ratio
        MEM	    Main memory bandwidth in MBytes/s
       NUMA	    Local and remote memory accesses
        QPI	    QPI Link Layer data
   RECOVERY	    Recovery duration
       SBOX	    Ring Transfer bandwidth
   TLB_DATA	    L2 data TLB miss rate/ratio
  TLB_INSTR	    L1 Instruction TLB miss rate/ratio
       UOPS	    UOPs execution info
  UOPS_EXEC	    UOPs execution
 UOPS_ISSUE	    UOPs issueing
UOPS_RETIRE	    UOPs retirement
CYCLE_ACTIVITY	Cycle Activities
\end{lstlisting}
\captionsetup{width=1\linewidth}\caption{List of PMC groups provided by Likwid tool on HCLServer02}
\label{pmc_groups}
\end{figure}
\vfill
\bibliographystyle{IEEEtran}
\bibliography{IEEEabrv,paper}

\end{document}